\documentclass[a4paper,11pt]{article}
\usepackage{jcappub} 
\usepackage{lineno}
\usepackage{color}
\usepackage{colortbl} 
\usepackage[dvipsnames]{xcolor}
\usepackage{amsmath, bm, bbm}
\usepackage{graphicx}
\usepackage{soul}

\usepackage{natbib}

\title{\boldmath Galactic Science with the LiteBIRD satellite: Spectral characterization of diffuse Galactic polarized emission at the angular power spectrum level}

\author[1]{S.\,Vinzl,}
\author[1]{J.\,Aumont,}
\author[2]{L.\,Vacher,}
\author[3,4]{R.\,T.\,Génova-Santos,}
\author[3]{D.\,Adak,}
\author[5,6]{A.\,Rizzieri,}
\author[7]{H.\,Akamatsu,}
\author[8]{E.\,Allys,}
\author[9]{A.\,Anand,}
\author[10,11,12]{C.\,Baccigalupi,}
% ...
\affiliation[1]{Univ Toulouse, CNES, CNRS, IRAP, Toulouse, France}
\affiliation[2]{Université Paris-Saclay, CNRS/IN2P3, IJCLab, 91405 Orsay, France}
\affiliation[3]{Instituto de Astrofísica de Canarias, E-38200 La Laguna, Tenerife, Canary Islands, Spain}
\affiliation[4]{Departamento de Astrofísica, Universidad de La Laguna (ULL), E-38206, La Laguna, Tenerife, Spain}
\affiliation[5]{Department of Physics, University of Oxford, Denys Wilkinson Building, Keble Road, Oxford OX1 3RH, UK}
\affiliation[6]{Université Paris Cité, CNRS, Astroparticule et Cosmologie, F-75013 Paris, France}
\affiliation[7]{International Center for Quantum-field Measurement Systems for Studies of the Universe and Particles (QUP), High Energy Accelerator Research Organization (KEK), Tsukuba, Ibaraki 305-0801, Japan}
\affiliation[8]{Laboratoire de Physique de l’École Normale Supérieure, ENS, Université PSL, CNRS, Sorbonne Université, Université de Paris, 75005 Paris, France}
\affiliation[9]{Dipartimento di Fisica, Università di Roma Tor Vergata, Via della Ricerca Scientifica, 1, 00133, Roma, Italy}
\affiliation[10]{International School for Advanced Studies (SISSA), Via Bonomea 265, 34136, Trieste, Italy}
\affiliation[11]{INFN Sezione di Trieste, via Valerio 2, 34127 Trieste, Italy}
\affiliation[12]{IFPU, Via Beirut, 2, 34151 Grignano, Trieste, Italy}
% ...

\emailAdd{samy.vinzl@utoulouse.fr}

\abstract
    {Detection of primordial $B$-mode polarization in the cosmic microwave background~(CMB) from tensor perturbations generated during inflation is a major scientific goal of future CMB missions. Its success will strongly depend on the characterization of polarized foregrounds, a challenge that the \textit{LiteBIRD} satellite aims to tackle with its $15$ frequency bands ranging from $40$ to $402$~GHz. In this work, we forecast the ability of \textit{LiteBIRD} to characterize polarized dust and synchrotron emission in the diffuse interstellar medium~(ISM), at the angular power spectrum level. From simulated \textit{LiteBIRD} intensity and polarization maps with different foreground complexities, we compute cross-frequency angular power spectra and fit them to dust and synchrotron spectral energy distributions, which are modeled by a modified black body and a power law, respectively. We find that \textit{LiteBIRD} will be able to measure the dust temperature, dust and synchrotron spectral indices and spatial correlation with dispersions as low as $\sigma(T_{\rm d}) \sim 0.2$~K, $\sigma(\beta_{\rm d}) \sim 0.006$, $\sigma(\beta_{\rm s}) \sim 0.04$ and $\sigma(\rho) \sim 10^{-2}$, as well as to detect and quantify deviations from the proposed parametric model due to variations of the emission properties in the three dimensions of our Galaxy. Additionally, \textit{LiteBIRD} is likely to rule out the power-law model of polarized foreground angular power spectra suggested by \textit{Planck} data. It will also be able to detect differences in the values of $\beta_{\rm d}$, $T_{\rm d}$, and $\beta_{\rm s}$ between $E$ modes, $B$ modes, and intensity in the diffuse ISM for the first time,  highlighting the joint variations of the physical conditions and the magnetic field structure across the Galaxy. We conclude that in addition to detailed studies of CMB polarization, \textit{LiteBIRD} will open a new window onto the physical conditions governing the ISM of the Milky Way.}


\begin{document}
\maketitle
\flushbottom


\section{Introduction}

Over the last three decades, the cosmic microwave background~(CMB) has proven to be a powerful probe of the early and late Universe. Measurements of its anisotropies using three space missions, \textit{COBE}~\cite{Smoot1992}, \textit{WMAP}~\cite{Bennett2013} and \textit{Planck}~\cite{Planck2018_IV} with increasing sensitivity as well as plenty of ground-based and balloon-borne experiments, e.g., \textit{ACT}~\cite{Naess2025}, \textit{SPT}~\cite{Camphuis2026}, have extensively advanced our understanding of the Universe. \par


Contemporary efforts are focusing on the study of the CMB polarization, arising from Thomson scattering of photons by electrons in the primordial plasma at the last scattering surface in the presence of an incident radiation field with a quadrupolar pattern. The observed CMB polarization also contains contributions from the late Universe through Compton and Thomson scattering during reionization, as well as the lensing of polarized CMB photons by high-redshift galaxy clusters, distorting the geometry of the originally produced polarization patterns~\cite{Seljak2000, Zaldarriaga1998}. CMB polarization can be decomposed into $E$ modes and $B$ modes, quantifying the contribution of curl-free and divergence-free polarized signal, respectively~\cite{Zaldarriaga1997, Kamionkowski1997b}. While scalar density fluctuations at the time of recombination can only produce $E$-mode polarization, the presence of primordial $B$ modes in the CMB is predicted to have a different origin. In $\Lambda$CDM cosmology, they can indeed only arise in the early Universe through tensor perturbations generated during inflation, an accelerated expansion phase occurring in the first fractions of a second of existence of the Universe~\cite{Brout1978, Starobinsky1980, Guth1981}, distorting the spherical symmetry of scalar perturbations and therefore producing divergence-free polarization patterns~\cite{Polnarev1985, Kamionkowski1997a, Seljak1997}. The ratio of amplitudes of these tensor and scalar perturbations is denoted as the tensor-to-scalar ratio, $r$, and is directly linked to the energy scale at which inflation occurred~\cite{Zaldarriaga2004}. Detecting the primordial $B$ modes or putting tighter constraints on $r$ is the main objective of current and future CMB experiments such as \textit{BICEP/Keck}~\cite{Tristram2022} and the \textit{Simons Observatory}~(\textit{SO})~\cite{Ade2019}. However, measuring the tensor-to-scalar ratio requires unprecedented sensitivity on angular scales $\gtrsim 1^\circ$~\cite{Kamionkowski2016}. Consequently, the aforementioned ground-based experiments, which cannot access the largest angular scales, may be limited by their restricted sky coverage. \par

However, the strongest limiting factor in such studies is undeniably our capacity to separate the faint primordial CMB polarization from the Galactic polarized signal~\cite{Planck2018_I, Planck2018_IV, Planck2018_XI}. Indeed, at frequencies $\gtrsim 100$~GHz, the microwave sky is dominated by thermal emission from dust grains heated by the interstellar radiation field~\cite{Krachmalnicoff2016}. This emission is strongly polarized due to the elongated shape of cold dust grains rotating around their minor axes, which are preferentially aligned with the local Galactic magnetic field~(GMF) lines, yielding polarization fractions up to $\sim 20 \, \%$~\cite{Planck2015_XIX, Planck2018_XII}. Synchrotron radiation originating from cosmic ray electrons spiraling around the GMF lines is the dominant source of polarization at frequencies~$\lesssim~100$~GHz. It can reach polarization fractions up to $\sim 10 \, \%$ at degree scale~\cite{Kogut2007}, and can therefore be a strong contaminant in the quest for primordial $B$ modes. Free-free emission, anomalous microwave emission~(AME)~\cite{Leitch1997, Draine1998a, Draine1998b}, CO spectral lines and the cosmic infrared background are also a strong source of contamination for observations of CMB temperature anisotropies, but are known to have negligible polarization levels~\cite{Planck2018_IV, Macellari2011, Herman2023, Genova-Santos2015, Tramonte2023, Gonzalez-Gonzalez2025, Feng2020}. \par

Foreground emission forces CMB experiments to include detectors covering a wide range of frequencies, with their precise characterization being of prime importance to separate them from the CMB radiation. The most recent space-based CMB experiment, \textit{Planck}, enabled detailed studies of these radiation processes (e.g., refs.~\cite{Planck2013_XIII, Planck2015_X, Planck2015_XXV}), including the nature of AME~\cite{Planck2014_XV}, zodiacal emission from interplanetary dust~\cite{Planck2013_XIV}, Galactic cold dust clumps~\cite{Planck2015_XXVIII}, interstellar dust composition~\cite{Planck2015_XXII}, GMF~\cite{Planck2016_XXXII, Planck2016_XXXV, Planck2016_XLII}, grain alignment~\cite{Planck2018_XII}, turbulence in the interstellar medium (ISM)~\cite{Planck2015_XX}, and more. \par

\textit{LiteBIRD} (the Lite (Light) satellite for the studies of $B$-mode polarization and Inflation from cosmic background Radiation Detection) represents the next generation of space-based CMB experiments, with the primary goal of placing stringent constraints on inflationary gravitational waves by measuring the tensor-to-scalar ratio with unprecedented accuracy $\sigma(r) \lesssim 10^{-3}$~\cite{PTEP2023}. By scanning the sky polarization from large to small angular scales in $15$ frequency bands ranging from $40$ to $402$~GHz, it will enable accurate separation of the faint primordial $B$-mode signal from Galactic foregrounds. This will provide an opportunity for detailed studies of Galactic polarized radiation processes in addition to the main cosmology science case. Despite its poorer angular resolution, \textit{LiteBIRD} will benefit from its increased sensitivity and broader frequency coverage compared to \textit{Planck}, promising major advances in understanding Galactic emission processes and probing the inner structure of the ISM and the GMF. \par

In this work, we simulate \textit{LiteBIRD} observations to forecast its ability to constrain the physical parameters of the two mechanisms of diffuse Galactic polarized emission: synchrotron and thermal dust. We use a power spectrum analysis to assess the extent to which \textit{LiteBIRD} will provide tight constraints on the dust and synchrotron spectral energy distributions (SEDs) and on their spatial correlation at intermediate and high Galactic latitudes, particularly on large scales where, in addition to a wide frequency range, it will benefit from its privileged location in space, unlike ground-based experiments such as \textit{SO}. This approach is formally equivalent to performing the analysis in pixel space and studying mean properties across the sky. It has, for instance, been used by \textit{Planck}, which showed that the thermal dust emission power spectra at intermediate and high Galactic latitudes can be well described by power laws in multipole space. \textit{Planck} also provided measurements of their $E$/$B$ ratio and correlation with synchrotron radiation~\cite{Planck2016_XXX, Planck2018_XI}. Similar forecasts are presented in ref.~\cite{Hensley2022} for \textit{SO}, this framework being further motivated by power-spectrum-based component separation methods in the quest for primordial $B$ modes. \par

We expect that the parameters describing the dust and synchrotron SEDs differ between $E$ and $B$ modes due to the spectral mixing of different emitting regions in the sky, which we call $E$/$B$-discrepancy in this paper. Such an effect, also referred to as the ``frequency dependence of the $E$/$B$ ratio''~\cite{Vacher2023, Ritacco2023}, is possible only in the presence of co-variation of the physical parameters and of the magnetic field structure, and has been detected at the map level in \textit{Planck} data along complex lines of sight~\cite{Pelgrims2021}. It is linked to but differs from the \emph{decorrelation}~\cite{Planck2016_XXX, Planck2017_L}, happening due to the spatial variations of the spectral parameters on the sky. For the same reason, we also expect the spectral parameters to differ between temperature and polarization, hereafter $T$/$P$-discrepancy. In this article, we assess the ability of \textit{LiteBIRD} to detect these effects for the first time, opening a new window on the understanding of Galactic polarized emission with microwave data. We show that these discrepancies can be explained and modeled by the moment expansion formalism~\cite{Chluba2017}, taking into account SED distortions induced by the averaging of SEDs with spatially varying spectral parameters along the line of sight, across the instrumental beam, and in harmonic space. \par

This paper is organized as follows. In section~\ref{sec:sims}, we first present how we produce accurate simulations of future \textit{LiteBIRD} observations with different Galactic emission models, presenting various levels of complexity. In section~\ref{sec:methods}, we then review the methods behind the power spectrum analysis and how we infer the parameters characterizing thermal dust and synchrotron radiation, before presenting our forecasts and interpreting them using moment expansion in section~\ref{sec:results}. Finally, we summarize and discuss our main results in section~\ref{sec:conclusion}.

\section{Sky simulations} \label{sec:sims}

In order to accurately forecast \textit{LiteBIRD} constraining power to characterize Galactic polarized emission, we produce different sets of sky simulations with different complexities. Every set of simulations is composed of a Galactic emission model containing thermal dust and synchrotron radiation, Gaussian realizations of the CMB, and end-to-end noise simulations taking into account \textit{LiteBIRD} instrumental characteristics. We smooth the sky maps with Gaussian beams for the respective channels listed in table~\ref{tab:baseline} before adding noise. We simulate $22$ sets of Stokes $I$, $Q$ and $U$ maps corresponding to each of the \textit{LiteBIRD} frequency channels at a \texttt{HEALPix}\footnote{\url{https://healpix.sourceforge.io/}} resolution of $N_{\rm side} = 64$~\cite{Gorski2005, Zonca2019}, assuming delta-function bandpasses. In a more realistic scenario, bandpass integration must be accounted for during the data analysis to avoid mismodeling the foreground SEDs. In this work, we assume that it is perfectly known and corrected for, leaving the more realistic case for future work. We produce all our maps and perform our analysis in thermodynamic units ($\mu$K$_{\rm CMB}$).

\begin{table}[t]
    \centering
    \begin{tabular}{cccc}
        \hline\hline\noalign{\vskip 1pt}
        Telescope & Frequency & Sensitivity, $\sigma^{Q,U}_1$ & Beam, $\theta_{\rm FWHM}$ \\
        & [GHz] & [$\mu$K$_{\rm CMB}$ arcmin] & [arcmin] \\
        \hline
        LFT & 40 & 37.42 & 70.5 \\
        LFT & 50 & 33.46 & 58.5 \\
        LFT & 60 & 21.31 & 51.1 \\
        LFT & 68 & 19.91/31.77 & 41.6/47.1 \\
        LFT & 78 & 15.55/19.13 & 36.9/43.8 \\
        LFT & 89 & 12.28/28.77 & 33.0/41.5 \\
        LFT/MFT & 100 & 10.34/8.48 & 30.2/37.8 \\
        LFT/MFT & 119 & 7.69/5.70 & 26.3/33.6 \\
        LFT/MFT & 140 & 7.25/6.38 & 23.7/30.8 \\
        MFT & 166 & 5.57 & 28.9 \\
        MFT/HFT & 195 & 7.05/10.50 & 28.0/28.6 \\
        HFT & 235 & 10.79 & 24.7 \\
        HFT & 280 & 13.80 & 22.5 \\
        HFT & 337 & 21.95 & 20.9 \\
        HFT & 402 & 47.45 & 17.9 \\
        \hline\hline
    \end{tabular}
    \vspace{0.1cm}
    \caption{\textit{LiteBIRD} instrumental characteristics used in this work for the Low-Frequency Telescope~(LFT), the Mid-Frequency Telescope~(MFT), and the High-Frequency Telescope~(HFT)~\cite{PTEP2023}. In cases where a frequency band is shared between two telescopes or detector arrays, the corresponding beam sizes and sensitivities are listed on the same line.}
    \label{tab:baseline}
\end{table}


\subsection{Galactic emission} \label{sec:fg}

We construct Galactic emission maps based on the Python Sky Model (\texttt{PySM}\footnote{\url{https://github.com/galsci/pysm}})~\cite{Thorne2017, Zonca2021, Panexp2025}. As we focus mainly on polarization data, we model only thermal dust emission and synchrotron radiation. We produce maps with the three levels of spectral complexity recommended in ref.~\cite{Panexp2025}, and quoted in table~\ref{tab:foregrounds}.

\begin{table}[t]
    \centering
    \begin{tabular}{cccc}
        \hline\hline
        Complexity & Low & Medium & High \\
        \hline
        Dust & \texttt{d9} & \texttt{d10} & \texttt{d12} \\
        Synchrotron & \texttt{s4} & \texttt{s5} & \texttt{s7} \\
        \hline
        Nomenclature & \texttt{d9s4} & \texttt{d10s5} & \texttt{d12s7} \\
        \hline\hline
    \end{tabular}
    \vspace{0.1cm}
    \caption{Sets of \texttt{PySM} models for Galactic polarized emission with different complexities.}
    \label{tab:foregrounds}
\end{table}

\subsubsection{Thermal dust emission}

The simplest way to model the thermal dust SED is to use a modified black body~(MBB) for each pixel,
\begin{equation}
    S_{\nu, \rm d}(A_{\rm d}, \beta_{\rm d}, T_{\rm d}) = A_{\rm d} \varepsilon_{\rm d}(\nu, \beta_{\rm d}, T_{\rm d}) = A_{\rm d} \left( \frac{\nu}{\nu_{\rm d}} \right)^{\beta_{ \rm d} - 2} \frac{B_\nu(\nu, T_{\rm d})}{B_\nu(\nu_{\rm d}, T_{\rm d})},
    \label{eq:MBB}
\end{equation}
where $\varepsilon_{\rm d}(\nu, \beta_{\rm d}, T_{\rm d})$ describes the frequency dependence of the SED, $B_\nu$ is the Planck function, $T_{\rm d}$ the dust temperature, $\beta_{\rm d}$ the dust spectral index, and $A_{\rm d}$ the SED amplitude at the reference frequency $\nu_{\rm d}$. Equation~\eqref{eq:MBB} is given in antenna temperature units (e.g., $\mu$K$_{\rm RJ}$), but we convert it into units of thermodynamic CMB temperature, $\mu$K$_{\rm CMB}$, for all our analyses. The conversion factor is given by
\begin{equation}
    f(\nu) = \frac{(e^x - 1)^2}{x^2 e^x} \quad \text{with} \quad x = \frac{h\nu}{kT_0},
    \label{eq:conversion}
\end{equation}
where $h$ is the Planck constant, $k$ the Boltzmann constant, and $T_0 = 2.7255$~K is the mean CMB temperature~\cite{Fixsen2009}. In this work, we use \textit{LiteBIRD}'s highest frequency, $\nu_{\rm d} = 402$~GHz, as a reference, but we verified that this choice does not affect our results. \par

The MBB model of the thermal dust SED is compatible with our most accurate measurements in the millimeter wavelengths~\cite{Planck2018_XI}, and is used in the three input templates used in this work, listed below.

\paragraph*{\tt low complexity:} A simple MBB parametrization is implemented within the \texttt{d9} model of \texttt{PySM}, where the dust temperature is fixed to $T_{\rm d} = 19.6$~K and the dust spectral index to $\beta_{\rm d} = 1.48$ across the sky. $I$, $Q$ and $U$ amplitude maps at $353$~GHz are retrieved from the \textit{Planck} \texttt{GNILC} component separation algorithm~\cite{Remazeilles2011, Planck2018_IV}, and are extrapolated to other frequencies using eq.~\eqref{eq:MBB}, with $\beta_{\rm d}$ and $T_{\rm d}$ having the same value in every pixel. This model is however greatly oversimplified, as physical conditions such as temperature and grain composition vary on all scales throughout the three dimensions of our Galaxy~\cite{Ferriere2001, Paradis2009, Ysard2013}, leading to corresponding variations in the spectral parameters~\cite{Zelko2022, Liu2024, Pelgrims2021}.

\paragraph*{\tt medium complexity:} An increased complexity is obtained with the \texttt{d10} \texttt{PySM} model, where both the dust temperature and spectral index vary across the sky from one pixel to another, according to the results of the \texttt{GNILC} component separation algorithm applied to \textit{Planck} intensity data. Amongst other observational consequences, these spatial variations cause SED distortions with respect to eq.~\eqref{eq:MBB}, as the SED of each pixel is no longer a perfect MBB~\cite{Chluba2017, Vacher2022b}. Indeed, \texttt{GNILC} foreground templates are produced at a native resolution of $N_{\rm side} = 2048$. We then generate maps using \texttt{PySM} at this resolution and subsequently downgrade them to $N_{\rm side} = 64$. Consequently, the resulting sky maps contain pixel averaging across lines of sight (LOS), which is further amplified by convolution with the instrumental beam. As the same procedure is used for every foreground model described in this section, this last statement remains true whenever the values of the corresponding spectral parameters vary across the sky.

\paragraph*{\tt high complexity:} The most complex polarized dust emission model used in this work is \texttt{d12}, which introduces variations in spectral properties along the LOS through the superposition of six dust emission layers, each characterized by a different plane-of-sky magnetic field component, thereby mimicking the effects of LOS frequency decorrelation~\cite{Tassis2015, Pelgrims2021}. Similar to the \texttt{d10} model, each layer exhibits spatially varying spectral parameters from pixel to pixel~\cite{Martinez_Solaeche_2018}. The dust emission is therefore averaged over the three dimensions, both across and along LOS.

\subsubsection{Synchrotron radiation}

The simplest model to describe the synchrotron SED is a power law (PL) in each pixel,
\begin{equation}
    S_{\nu, \rm s}(A_{\rm s}, \beta_{\rm s}) = A_{\rm s} \varepsilon_{\rm s}(\nu, \beta_{\rm s}) = A_{\rm s} \left( \frac{\nu}{\nu_{\rm s}} \right)^{\beta_{\rm s}},
    \label{eq:PL}
\end{equation}
where $\varepsilon_{\rm s}(\nu, \beta_{\rm s})$ is the synchrotron frequency dependence, $\beta_{\rm s}$ the synchrotron spectral index and $A_{\rm s}$ is the amplitude of the SED at an arbitrary reference frequency, which we take to be \textit{LiteBIRD}'s lowest frequency band, $\nu_{\rm s} = 40$~GHz. Again, we verified that this choice has no impact on our results. This model is motivated by the fact that the energy distribution of cosmic ray electrons responsible for synchrotron radiation is typically characterized by a PL (e.g., ref.~\cite{Rybicki1979}). The synchrotron SED in eq.~\eqref{eq:PL} is given in antenna temperature units~($\mu$K$_{\rm RJ}$). In the remainder of this article, as we do for thermal dust emission, we express the synchrotron surface brightness in units of $\mu$K$_{\rm CMB}$ by applying the conversion factor defined in eq.~\eqref{eq:conversion}. \par

\paragraph*{\tt low complexity:} We use the \texttt{s4} model of \texttt{PySM} as our simplest synchrotron model. The emission templates are taken to be the Haslam $408$~MHz data map~\cite{Haslam1981, Haslam1982} reprocessed by ref.~\cite{Remazeilles2015} for intensity, and the \textit{WMAP} nine-year $23$~GHz $Q$ and $U$ maps~\cite{Bennett2013} for polarization. These templates are then extrapolated to other frequencies using the synchrotron SED model of eq.~\eqref{eq:PL}, assuming a fixed spectral index $\beta_{\rm s} = -3.1$ across the sky.

\paragraph*{\tt medium complexity:} As our intermediate complexity Galactic emission model, we use the \texttt{s5} model, in which the spectral index varies in each pixel of the sky. The input $\beta_{\rm s}$ template maps are derived by combining the Haslam $408$~MHz, \textit{WMAP} seven-year $23$~GHz~\cite{Miville-Deschenes2008}, and \textit{S-PASS} $2.3$~GHz~\cite{Krachmalnicoff2018} data. Again, we first generate synchrotron templates using \texttt{PySM} at $N_{\rm side} = 2048$ and then downgrade them to $N_{\rm side} = 64$. Therefore, the 2D spectral variations introduced in this model cause SED distortions in large pixels.

\paragraph*{\tt high complexity:} The most complex model contains the \texttt{s7} \texttt{PySM} maps, introducing curvature in the synchrotron SED through the curvature parameter $c_{\rm s}$,
\begin{equation}
    S_{\nu, \rm s}^{\rm curved}(A_{\rm s}, \beta_{\rm s}, c_{\rm s}) = A_{\rm s} \varepsilon_{\rm s}^{\rm curved}(\nu, \beta_{\rm s}, c_{\rm s}) = A_{\rm s} \left( \frac{\nu}{\nu_{\rm s}} \right)^{\beta_{\rm s} + c_{\rm s} \log{(\nu / \nu_{\rm s})}},
    \label{eq:PL_curved}
\end{equation}
where the $A_{\rm s}$ and $\beta_{\rm s}$ template maps are identical to those of the \texttt{s5} model, and $c_{\rm s}$ is also varying across the sky with typical values of $c_{\rm s} \sim -0.05$, based on the intensity measurements from the \textit{ARCADE} balloon-borne experiment~\cite{Kogut2012}. The flattening of the PL at low frequencies, modeled by $c_{\rm s} < 0$, is thought to originate from the various interactions of the relativistic particles propagating through the ISM~\cite{Kogut2012, Orlando2013}, such as radiative losses and Faraday rotation. Conversely, a positive curvature can also result from the superposition of multiple components along the LOS~\cite{Weymann-Despres2026}. The \textit{QUIJOTE} experiment~\cite{Rubino-Martin2023} has found tentative evidence for curvature in the synchrotron polarized spectrum, with a uniform parameter $c_{\rm s} = -0.797 \pm 0.0012$, but the data lack sufficient statistical power to claim an unambiguous detection~\cite{delaHoz2023}. \par

In the foreground simulations presented above, the spectral parameters are assumed to be identical in intensity and polarization, which may be an oversimplifying assumption. However, during the analysis, they are fitted independently for intensity, $E$ modes and $B$ modes, see section~\ref{sec:fits}.

\subsection{Cosmic microwave background} \label{sec:cmb}

We simulate Gaussian random realizations of CMB $I$, $Q$ and $U$ maps from $TT$, $EE$ and $BB$ CMB power spectra using the \texttt{synfast} function of \texttt{HEALPix}. These spectra are generated with the Code for Anisotropies in the Microwave Background (\texttt{CAMB}\footnote{\url{https://camb.info/}})~\cite{Lewis2000} using the \textit{Planck} 2018 best-fit cosmological parameters~\cite{Planck2018_I, Planck2018_VI}, assuming $r=0$. Introducing non-zero tensor modes would not significantly affect the conclusions of this paper, as foreground emission largely dominates over the primordial $B$-mode signal. We neglect the non-Gaussian properties of lensing $B$ modes, as this study focuses solely on the Galactic foreground signal.

\subsection{Instrumental noise} \label{sec:noise}

We add to our $I$, $Q$ and $U$ sky maps the \textit{LiteBIRD} end-to-end noise simulations released in ref.~\cite{Bortolami2025}. They consist of a combination of Gaussian white noise with zero mean and 
$1/f$ noise with a knee frequency $f_{\rm knee} = 30$~mHz. In polarization, the latter is strongly reduced by the continuously rotating half-wave plates~\cite{PTEP2023}, yielding Gaussian white noise with a standard deviation given by $\sigma^{Q,U}_{\rm pix}(\nu) = \sigma^{Q,U}_1(\nu) / \sqrt{\Omega_{\rm pix}}$, where $\Omega_{\rm pix}$ is the pixel solid angle in arcmin$^2$, and $\sigma^{Q,U}_1(\nu)$ is the sensitivity at frequency $\nu$ for a resolution element of $1$ arcmin$^2$. These values, along with the associated beam size for each frequency channel, are listed in table~\ref{tab:baseline}. \par


To avoid noise bias in the power spectra of data at the same frequency, we also produce sets of maps corresponding to half missions, with statistically independent noise whose level is increased by a factor of $\sqrt{2}$. These maps are then combined to compute cross-spectra that are free from instrumental noise bias (see section~\ref{sec:cross}). \par



To compare the results obtained for \textit{LiteBIRD} sensitivity with those for \textit{Planck}, we also simulate the sky maps at \textit{Planck} frequency channels. The noise maps used are taken from the \textit{Planck} full-focal plane~(FFP10) simulations~\cite{Planck2018_III}.


\section{Power spectrum analysis} \label{sec:methods}

We compute the cross-frequency angular power spectra from the $I$, $Q$ and $U$ maps between \textit{LiteBIRD} $N_{\rm freq} = 22$ frequency bands, as well as the cross-power spectra between the two half mission (HM) map subsets at the same frequency. This yields a total of $N_{\rm cross} = N_{\rm freq} (N_{\rm freq}+1)/2 = 253$ power spectra (see section~\ref{sec:cross}), $231$ of which involve different frequencies, and $22$ of which are computed from the two HM maps of the same frequency. \par

\subsection{Galactic masks} \label{sec:masks}

In this work, we focus on the dust and synchrotron properties in the diffuse ISM, from intermediate to high Galactic latitudes. To exclude the Galactic plane from the analysis, where the emission properties differ from other regions~\cite{Planck2013_XI, Planck2014_XIV, Planck2015_XIX, Planck2018_XII}, we apply to the simulated maps the Galactic masks released in the  \textit{Planck} 2015 data analysis~\cite{Planck2015_I}. Forecasts for \textit{LiteBIRD} science cases at low Galactic latitudes are left for future work. \par

\begin{figure}[t]
    \centering
    \includegraphics[width=0.6\linewidth]{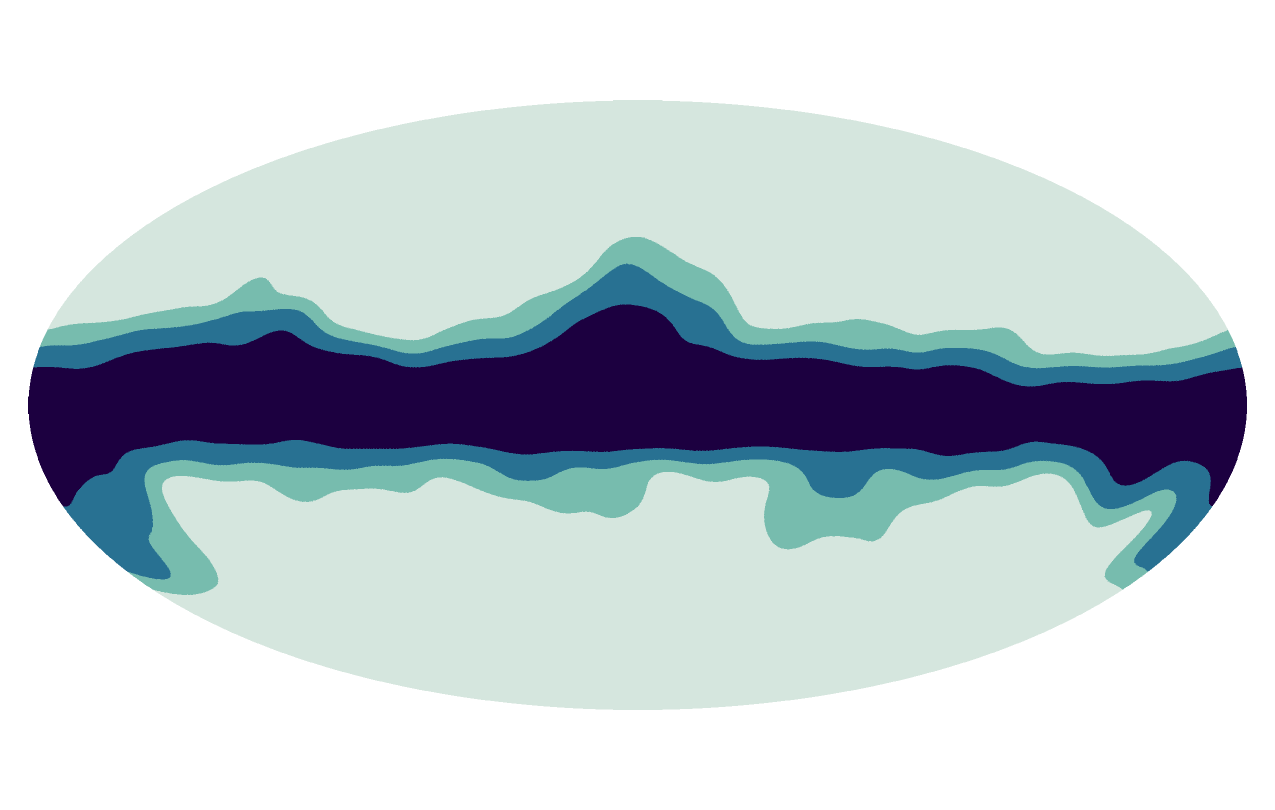}
    \caption{Binary Galactic masks used in this study. The masked regions are shown in dark blue, cyan, and green, corresponding to observed sky fractions of $f_{\rm sky} = 0.6$, $0.7$, and $0.8$, respectively~\cite{Planck2015_I}.}
    \label{fig:masks}
\end{figure}

We use the \texttt{GAL060}, \texttt{GAL070} and \texttt{GAL080} masks, leaving unmasked sky fractions of $f_{\rm sky} = 0.6$, $f_{\rm sky} = 0.7$, and $f_{\rm sky} = 0.8$, respectively. The binary versions of these masks are shown in figure~\ref{fig:masks}. In section~\ref{sec:results}, we present the results obtained with $f_{\rm sky} = 0.7$, which we choose to be the baseline of this paper. A comparison with the results obtained with $f_{\rm sky} = 0.6$ and $f_{\rm sky} = 0.8$ is presented in appendix~\ref{app:fsky_comparison}. \par

\subsection{Cross-spectra estimation} \label{sec:cross}

We compute all the cross-frequency angular power spectra $\mathcal{C}_\ell^{XX}(\nu_i \times \nu_j) = \left\langle a_{\ell m}^X(\nu_i) a_{\ell m}^{X*}(\nu_j) \right\rangle$ between our simulated maps at frequency $\nu_i$ and $\nu_j$ using the \texttt{NaMaster}\footnote{\url{https://github.com/LSSTDESC/NaMaster}} Python package~\cite{Alonso2019}. In the above expression, $X$ stands for temperature $T$, $E$-mode, or $B$-mode polarization. As mentioned in section~\ref{sec:noise}, we do not compute directly the $\mathcal{C}_\ell^{XX}(\nu_i^{\rm FM} \times \nu_i^{\rm FM})$ auto-spectra using the full mission (FM) maps. Instead, we compute the cross-spectra $\mathcal{C}_\ell^{XX}(\nu_i^{\rm HM1} \times \nu_i^{\rm HM2})$ between half mission (HM) maps at frequency $\nu_i$. This makes it possible to mitigate the noise bias in the auto-spectra due to the instrumental noise auto-correlation, which may be difficult to model accurately in real observations. \par

Computing power spectra on a cut sky leads to mixing between $E$ and $B$ modes, increasing the variance of the power spectra. To mitigate this effect, we use the \texttt{NaMaster} purification function to purify $E$ and $B$ modes when computing $EE$ and $BB$ power spectra, respectively. However, when applying purification, the mask needs to be apodized to avoid bias in the estimated power spectra. In this work, we choose the \texttt{C1} apodization type of \texttt{NaMaster} with a $10^\circ$ apodization scale. To verify that this choice leads to unbiased estimates of the power spectra, we produced two validation datasets: one containing only the CMB signal, and another corresponding to power-law power spectra closer to those of the foregrounds. By computing the power spectra of $1000$ such simulations, we verified that the resulting estimates are unbiased in both cases over the full range of angular scales.

Additionally, we correct for the finite resolution effects by dividing the computed cross-power spectra by the product of the beam transfer functions at frequencies $\nu_i$ and $\nu_j$, given by
\begin{equation}
    B_\ell(\nu) =
    \begin{cases}
        \exp{\left[ -\frac{\ell(\ell+1)}{2} \frac{\theta_{\rm FWHM}^2(\nu)}{8 \log{2}} \right]} & \text{for spin-0 fields} \\
        \exp{\left[ -\frac{\ell(\ell+1) - 4}{2} \frac{\theta_{\rm FWHM}^2(\nu)}{8 \log{2}} \right]} & \text{for spin-2 fields},
    \end{cases}
\end{equation}
see e.g.~\cite{Challinor2000}. This requires the assumption that the beams are perfectly Gaussian and known during the data analysis. \par

Conservatively, the cross-power spectra are estimated at multipoles $\ell < 2 N_{\rm side} = 128$ using \texttt{NaMaster}~\cite{Alonso2019}. As a computationally manageable case, they are binned into $N_{\rm bin} = 12$ bandpowers of constant width $\Delta\ell = 10$, while still providing a sufficiently large number of bins to study the foregrounds. In the following, $\ell$ thus denotes the multipole bin of size $\Delta \ell = 10$ centered on $\ell$.

\subsection{Covariance matrix estimation} \label{sec:covmat}

The cross-frequency angular power spectra computed in section~\ref{sec:cross} are highly correlated in each multipole bin $\ell$. Additionally, because of the applied Galactic mask, we expect each cross-frequency angular power spectrum to show correlations between adjacent multipoles~\cite{Tristram2005}. We therefore need an accurate estimate of the covariance matrix of observations $\Xi_{\ell\ell'}^{XX}(\nu_i \times \nu_j, \nu_k \times \nu_l) = \mathrm{cov} [\mathcal{C}_\ell^{XX}(\nu_i \times \nu_j), \mathcal{C}_{\ell'}^{XX}(\nu_k \times \nu_l)]$ before comparing them with any model.  This covariance matrix should account not only for the contributions of the instrumental noise, but also for the cosmic variance arising from the limited number of independent modes on the largest scales. Furthermore, cross-correlations between the noise, CMB, and foregrounds make significant contributions to the angular power spectra covariance matrix and must therefore be accurately accounted for. For instance, ref.~\cite{Liu2025} showed, in the case of \textit{SO}, that an inaccurate modeling of the dust contribution to the covariance matrix can strongly degrade the quality of the fits. \par

To model this complexity, we perform an analytical estimate of the cross-frequency angular power spectra covariance matrix using the \texttt{NaMaster} covariance facility introduced in ref.~\cite{Garcia-Garcia2019}. This has the advantages of being both faster, and more accurate than estimating it from Monte Carlo simulations of CMB and instrumental noise. Indeed, a Monte-Carlo-based estimate would require at least $(N_{\rm bin} \cdot N_{\rm cross})^2 \sim 10^7$ simulations to achieve sufficient accuracy, which would be computationally prohibitive. We construct the covariance matrix from a fiducial power spectrum model containing CMB, foregrounds and instrumental noise,
\begin{equation}
    \tilde{\mathcal{C}}_\ell^{XX}(\nu_i \times \nu_j) = \tilde{\mathcal{C}}_{\ell,\rm CMB}^{XX} + \tilde{\mathcal{C}}_{\ell,\rm fg}^{XX}(\nu_i \times \nu_j) + \tilde{N}_\ell(\nu_i \times \nu_j),
    \label{eq:covmat}
\end{equation}
where the tilde denotes the mode-coupled pseudo-$\mathcal{C}_\ell$ divided by $f_{\rm sky}$, following the improved Narrow Kernel Approximation (iNKA)~\cite{Nicola2021}. The noise term is given by
\begin{equation}
    N_\ell(\nu_i^\alpha \times \nu_j^\beta) = \frac{\delta_{ij} g_{\alpha\beta}}{B_\ell^2(\nu_i)} \left[ N_{\rm white}(\nu_i) + N_{\rm corr}(\nu_i) \left( \frac{\ell}{\ell_{\rm knee}} \right)^{\alpha_{\rm knee}} \right],
    \label{eq:noise}
\end{equation}
where $\delta_{ij}$ is the Kronecker delta, $(\alpha, \beta) \in \{{\rm FM}, {\rm HM}\}$ denote if maps $i$ and $j$ are computed from FM or HM datasets, $N_{\rm white}(\nu_i)$ is the white noise level at frequency $\nu_i$, and the correlated noise is modeled by a power law of amplitude $N_{\rm corr}(\nu_i)$, knee multipole $\ell_{\rm knee}$, and index $\alpha_{\rm knee}$. These four quantities are fitted by averaging the angular power spectra of the input noise simulations described in section~\ref{sec:noise}. The term $g_{\alpha\beta} = \delta_{\alpha\beta}(1+\delta_{\alpha{\rm HM}})$ ensures that no noise contribution is computed for pairs of maps at the same frequency presenting uncorrelated noise. Indeed, $g_{\alpha\beta}=0$ if the two maps are different, $g_{\alpha\beta}=1$ if they are computed from the same FM dataset, and $g_{\alpha\beta}=2$ if they correspond to the same HM map. The latter arises because the noise level in each HM map is higher by a factor of $\sqrt{2}$ (see section~\ref{sec:noise}), which translates into a factor of $2$ at the power spectrum level. It should be noted that, even though we do not compute cross-power spectra from identical maps (see section~\ref{sec:cross}), the calculation of the covariance matrix nonetheless involves such terms. For instance, the computation of $\Xi_{\ell \ell'}^{BB}(100 \times 119, 119 \times 140)$ involves a term in $\mathcal{C}_\ell^{BB}(119 \times 119)$, which is in this case computed from the same FM dataset. \par

The noise terms of eq.~\eqref{eq:noise} and the theoretical CMB power spectra are coupled to the Galactic mask using \texttt{NaMaster} before being injected into eq.~\eqref{eq:covmat}. The foreground terms are directly obtained by computing the pseudo-$\mathcal{C}_\ell$ estimators of the different cross-frequency angular power spectra from the \texttt{PySM} models presented in section~\ref{sec:fg}. In a more realistic scenario, however, the foreground terms should be treated as unknown. In this case, a better methodology would be, for instance, to initialize the fits using a fiducial covariance matrix and then iteratively update the covariance matrix after each iteration. Another alternative would be to directly inject the total pseudo-power spectra computed from the data into eq.~\eqref{eq:covmat}, thereby avoiding the need to evaluate an instrumental noise model. A detailed study of these different estimation methods and their impact on the fits will be presented in future work. \par

In our simulations, Galactic foregrounds are not stochastically modeled but are directly extracted from \texttt{PySM} models. Therefore, the covariance should not include any foreground--foreground auto-correlation terms. However, cross-correlations between the foregrounds, noise, and the CMB are still present. We subtract foreground auto-correlations from the computed covariance matrix by evaluating the covariance matrix that would be obtained if only Galactic emission were present. This procedure accounts for the non-Gaussian spatial distributions of polarized thermal dust and synchrotron emission in the Galaxy. It should be noted that, in the case where the covariance matrix is computed directly from the data maps, a fiducial model of Galactic emission is still required for this purpose. \par

\begin{figure}[t]
    \centering
    \includegraphics[width=0.6\linewidth]{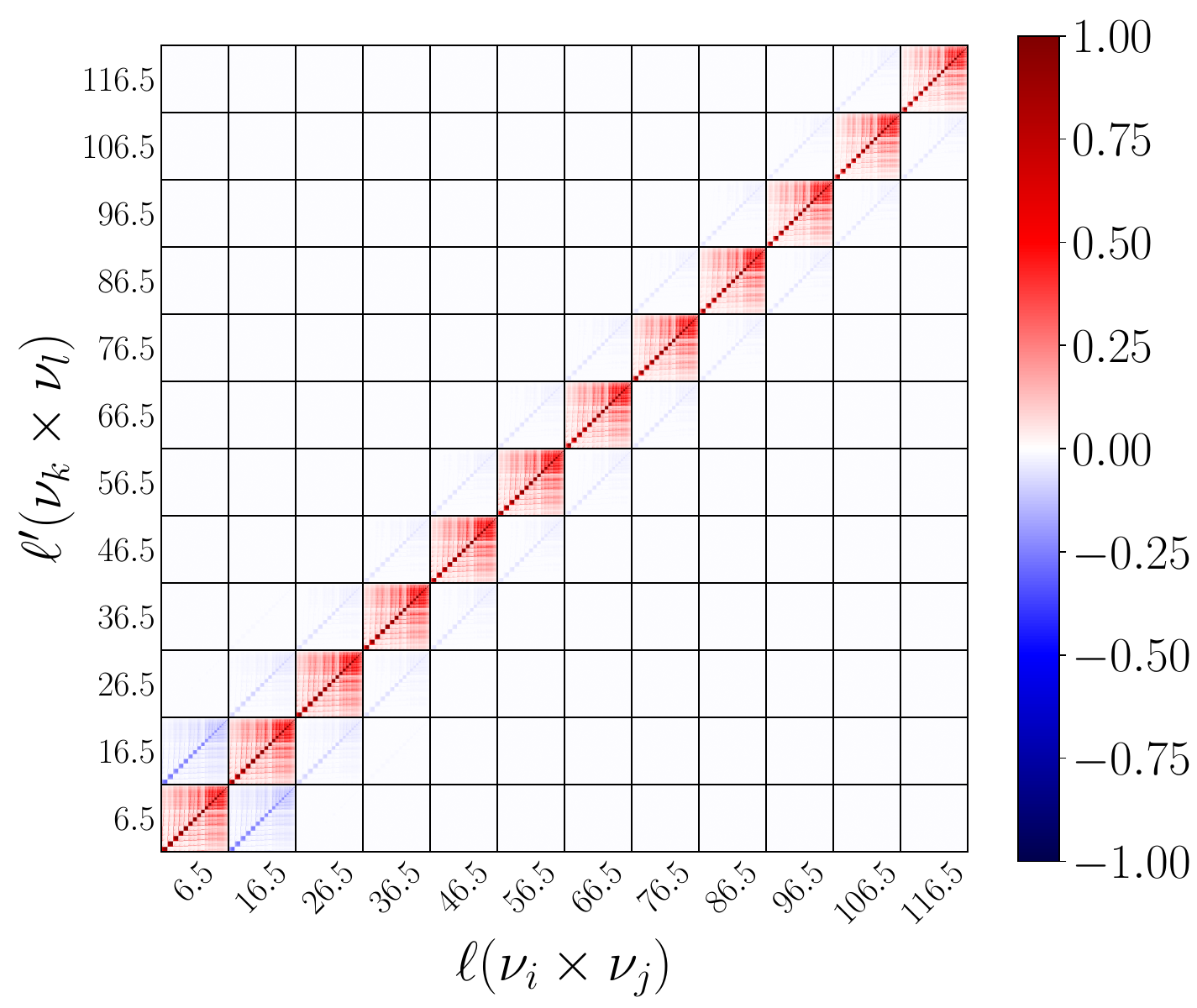}
    \caption{Correlation matrix $\Xi_{\ell\ell'} / \sqrt{\Xi_{\ell\ell} \Xi_{\ell'\ell'}}$ of the $B$-mode cross-frequency angular power spectra for the simplest Galactic emission model of table~\ref{tab:foregrounds} with $f_{\rm sky} = 0.7$, computed using \texttt{NaMaster}. The $N_{\rm bin} = 12$ multipole bins from $\ell = 6.5$ to $\ell = 116.5$ are represented by the black squares, each of them containing the ordered $N_{\rm cross} = 253$ cross-frequency angular power spectra.}
    \label{fig:covmat}
\end{figure}

The estimated angular power spectra covariance matrix is illustrated in figure~\ref{fig:covmat}, showing the computed correlation matrix $\Xi_{\ell\ell'} / \sqrt{\Xi_{\ell\ell} \Xi_{\ell'\ell'}}$ of the $B$-mode power spectra\footnote{In figure~\ref{fig:covmat}, the correlations between all the cross-frequency angular power spectra are shown between each couple of multipoles $(\ell, \ell')$ in the black squares. Power spectra are ordered from lowest to highest frequencies, namely $\nu_i \times \nu_j \in \{ 40 \times 40, 40 \times 50, \dots, 40 \times 402, 50 \times 50, \dots, 50 \times 402, \dots, 337 \times 402, 402 \times 402 \}$~GHz.} for our simplest Galactic emission model of table~\ref{tab:foregrounds} with $f_{\rm sky} = 0.7$. The obtained correlation matrix is in visual agreement with the simulation-based estimation of ref.~\cite{Vacher2022a}, see their figure~\href{https://www.aanda.org/articles/aa/pdf/2022/04/aa42664-21.pdf\#page=7}{2}. Using a Monte Carlo approach, the retrieved covariance matrix is essentially a noisy version of the one obtained with \texttt{NaMaster}. With our analytical estimate, we can therefore model with high accuracy all correlations between the different cross-spectra arising from auto- and cross-correlations among instrumental noise, the CMB, and Galactic emission.\footnote{We also performed tests using toy simulations with a limited number of frequency bands to ensure that our analytical estimate is compatible with the Monte Carlo approach. A detailed analysis of the effects of the different covariance estimates is left for future work.} \par

\subsection{Fitting procedure} \label{sec:fits}

We fit the computed cross-frequency angular power spectra in each multipole bin with a parametric model using the \texttt{mpfit}\footnote{\url{https://github.com/segasai/astrolibpy/tree/master/mpfit}} $\chi^2$ minimization package~\cite{Markwardt2009}. We consider the parametric model introduced in ref.~\cite{Choi2015} and also used in ref.~\cite{Planck2018_XI, Hensley2022},
\begin{equation}
    \begin{split}
        \mathcal{D}_\ell^{XX}(\nu_i \times \nu_j) =&~\mathcal{D}_{\ell,\rm CMB}^{XX}(\nu_i \times \nu_j) \\
        &\!\! + A_{\ell, \rm d}^{XX} \varepsilon_{\ell, \rm d}^{XX}(\nu_i) \varepsilon_{\ell,\rm d}^{XX}(\nu_j) + A_{\ell, \rm s}^{XX} \varepsilon_{\ell,\rm s}^{XX}(\nu_i) \varepsilon_{\ell,\rm s}^{XX}(\nu_j) \\
        &\!\! + \rho_\ell^{XX} \sqrt{A_{\ell, \rm d}^{XX} A_{\ell, \rm s}^{XX}} \left[ \varepsilon_{\ell,\rm d}^{XX}(\nu_i) \varepsilon_{\ell,\rm s}^{XX}(\nu_j) + \varepsilon_{\ell,\rm s}^{XX}(\nu_i) \varepsilon_{\ell,\rm d}^{XX}(\nu_j) \right],
    \end{split}
    \label{eq:model}
\end{equation}
where we have defined the usual rescaling of the cross-frequency angular power spectra
\begin{equation}
    \mathcal{D}_\ell^{XX}(\nu_i \times \nu_j) = \frac{\ell(\ell+1)}{2\pi} \mathcal{C}_\ell^{XX}(\nu_i \times \nu_j).
\end{equation}
In eq.~\eqref{eq:model}, $A_{\ell, \rm d}^{XX}$ and $A_{\ell, \rm s}^{XX}$ are the amplitudes of the dust and synchrotron power spectra at frequencies $\nu_{\rm d} = 402$~GHz and $\nu_{\rm s} = 40$~GHz, respectively, and $\rho_\ell^{XX}$ is the correlation coefficient between dust and synchrotron emission at multipole $\ell$. In this expression, $\varepsilon_{\ell,\rm d}^{XX}(\nu) = \varepsilon_{\rm d}^{XX}(\nu, \beta_{\rm d}(\ell), T_{\rm d}(\ell))$ and $\varepsilon_{\ell,\rm s}^{XX}(\nu) = \varepsilon_{\rm s}^{XX}(\nu, \beta_{\rm s}(\ell))$ are the frequency dependencies of dust and synchrotron emission defined in eqs.~\eqref{eq:MBB} and \eqref{eq:PL}, respectively, with $\ell$-dependent spectral parameters. In our Galactic emission models, the dust and synchrotron spectral indices are assumed to be the same in intensity and polarization. Nonetheless, $T_{\rm d}(\ell)$, $\beta_{\rm d}(\ell)$ and $\beta_{\rm s}(\ell)$ may differ among the $TT$, $EE$, and $BB$ power spectra and among multipoles due to the spectral mixing from different regions of the sky~\cite{Vacher2023}. In our simplest foreground model, \texttt{d9s4}, spectral parameters are constant across the sky (see section~\ref{sec:fg}). Therefore, we expect their values to be constant over multipole bins as well.

In this work, we discuss results obtained with two models for the power spectrum amplitudes. The first, proposed by ref.~\cite{Planck2016_XXX}, scales the amplitude as a power law of $\ell$,
\begin{equation}
    A_{\ell, \rm c}^{XX} = A_{0, \rm c}^{XX} \left( \frac{\ell}{\ell_0} \right)^{\alpha_{\rm c}},
    \label{eq:PL_spectra}
\end{equation}
where $\rm c$ stands for dust or synchrotron and $A_{0, \rm c}^{XX}$ is the amplitude at an arbitrary reference multipole $\ell_0$. In this case, all other parameters are assumed to be constant across multipoles. Equation~\eqref{eq:model} therefore becomes
\begin{equation}
    \begin{split}
        \mathcal{D}_\ell^{XX}(\nu_i \times \nu_j) =&~\mathcal{D}_{\ell,\rm CMB}^{XX}(\nu_i \times \nu_j) \\
        &\!\! + A_{0, \rm d}^{XX} \left( \frac{\ell}{\ell_0} \right)^{\alpha_{\rm d}} \varepsilon_{\rm d}^{XX}(\nu_i) \varepsilon_{\rm d}^{XX}(\nu_j) + A_{0, \rm s}^{XX} \left( \frac{\ell}{\ell_0} \right)^{\alpha_{\rm s}} \varepsilon_{\rm s}^{XX}(\nu_i) \varepsilon_{\rm s}^{XX}(\nu_j) \\
        &\!\! + \rho^{XX} \sqrt{A_{0, \rm d}^{XX} A_{0, \rm s}^{XX}} \left( \frac{\ell}{\ell_0} \right)^{\frac{\alpha_{\rm d}+\alpha_{\rm s}}{2}} \left[ \varepsilon_{\rm d}^{XX}(\nu_i) \varepsilon_{\rm s}^{XX}(\nu_j) + \varepsilon_{\rm s}^{XX}(\nu_i) \varepsilon_{\rm d}^{XX}(\nu_j) \right],
    \end{split}
    \label{eq:PL_model}
\end{equation}
which we refer to as the power-law model of the foreground power spectra. However, this parametric model may not remain adequate for future CMB experiments with enough sensitivity. For instance, it was shown that the power-law model might already be limiting for complex dust models in the case of \textit{SO}~\cite{Liu2025}. Furthermore, as discussed in section~\ref{sec:PL_spectra}, we witness that the power-law assumption breaks down in the case of \textit{LiteBIRD}, even with the simplest dust models. Therefore, following refs.~\cite{Planck2018_XI, Vacher2022a}, we also perform fits assuming independent amplitudes in each $\ell$-bin, thereby avoiding any shape dependency in the analysis. \par


In order to obtain an accurate estimation of the posterior distribution of the parameters for each Galactic emission model and for each Galactic mask (see section~\ref{sec:results}), we produce a set of $N_{\rm sim} = 250$ simulations for each case and fit them with eq.~\eqref{eq:model} by performing a Levenberg-Marquardt least-squares minimization using \texttt{mpfit}. That is to say, we minimize
\begin{equation}
    \chi^2 = \left( \bm{\mathcal{D}_{\ell,\rm data}^{XX}} - \bm{\mathcal{D}_{\ell,\rm model}^{XX}} \right)^\top \mathbf{\Xi}^{-1} \left( \bm{\mathcal{D}_{\ell,\rm data}^{XX}} - \bm{\mathcal{D}_{\ell,\rm model}^{XX}} \right),
    \label{eq:likelihood}
\end{equation}
where $\bm{\mathcal{D}_{\ell,\rm data}^{XX}}$ is the vector containing the simulated cross-frequency angular power spectra, $\bm{\mathcal{D}_{\ell,\rm model}^{XX}}$ is the vector containing the model, and $\mathbf{\Xi}$ is the covariance matrix of dimension $(N_{\rm bin} \cdot N_{\rm cross})^2$ computed in section~\ref{sec:covmat}. \par

As figure~\ref{fig:covmat} illustrates, the strongest correlations are the positive correlations between cross-frequencies at the same multipole (block-diagonal terms) and the anti-correlations between adjacent multipole bins (off-diagonal blocks). Therefore, we apply the following truncation of the covariance matrix: we infer the model parameters of each multipole $\ell$ by a joint fit to its adjacent multipoles. The first and last bins are retrieved together with the second and penultimate ones. In this way, we take into account the strongest correlations while considerably reducing the computing time. This approach is similar to that of ref.~\cite{Vacher2022a}, with the addition of correlations between adjacent multipole bins made possible by our analytical estimate of the covariance matrix.\footnote{Ref.~\cite{Vacher2022a} also takes into account the anti-correlations between the two first multipole bins, which are the strongest and therefore well approximated by Monte Carlo simulations.} \par

Six parameters per bin are allowed to vary during the fitting process: $A_{\ell, \rm d}^{XX}$, $T_{\rm d}$, $\beta_{\rm d}$, $A_{\ell, \rm s}^{XX}$, $\beta_{\rm s}$ and $\rho_\ell^{XX}$. We adopt wide flat priors for all parameters, restricting them to physically meaningful ranges: $T_{\rm d} \in [1,100]$~K, $\beta_{\rm d} \in [0,10]$, $\beta_{\rm s} \in [-10,0]$, $A_{\ell, \rm c}^{XX} \geq 0$~$\mu$K$_{\rm CMB}^2$, and $\rho_\ell^{XX} \in [-1,1]$. For each multipole bin $\ell$, the CMB spectrum $\mathcal{D}_{\ell,\rm CMB}^{XX}$ is fixed to its input value so that the CMB contributes only to the variance. Joint estimation of foreground and CMB components, which is the typical component separation problem, is beyond the scope of this paper and is left for future work. From the fits of our $N_{\rm sim}$ simulations, we compute the mean and standard deviation of the fitted parameters at each angular scale $\ell$, as well as their distribution using the Monte Carlo samples analysis Python package \texttt{GetDist}\footnote{\url{https://github.com/cmbant/getdist}}~\cite{Lewis2025}.

\section{Results} \label{sec:results}

We fit the cross-frequency angular power spectra computed in section~\ref{sec:cross} using the procedure detailed in section~\ref{sec:methods}, and in this section we present our results. In section~\ref{sec:PL_spectra}, we evaluate the ability of the power-law model of eq.~\eqref{eq:PL_model} to reproduce the computed power spectra. As it provides only poor fits, we adopt $\ell$-by-$\ell$ amplitudes in the remainder of the paper. In section~\ref{sec:model_comparison}, we compare the fits obtained for the different Galactic emission models for \textit{LiteBIRD} and \textit{Planck}. In section~\ref{sec:discrepancies}, we investigate how \textit{LiteBIRD} will enable the detection of differences in spectral parameters between $T$, $E$, and $B$ cross-frequency angular power spectra, which we refer to as $E$/$B$- and $T$/$P$-discrepancies. In section~\ref{sec:mom}, we use the moment expansion formalism to interpret these discrepancies.

\subsection{Power-law model of the foreground power spectra} \label{sec:PL_spectra}

\begin{figure}[t]
    \centering
    \includegraphics[width=\linewidth]{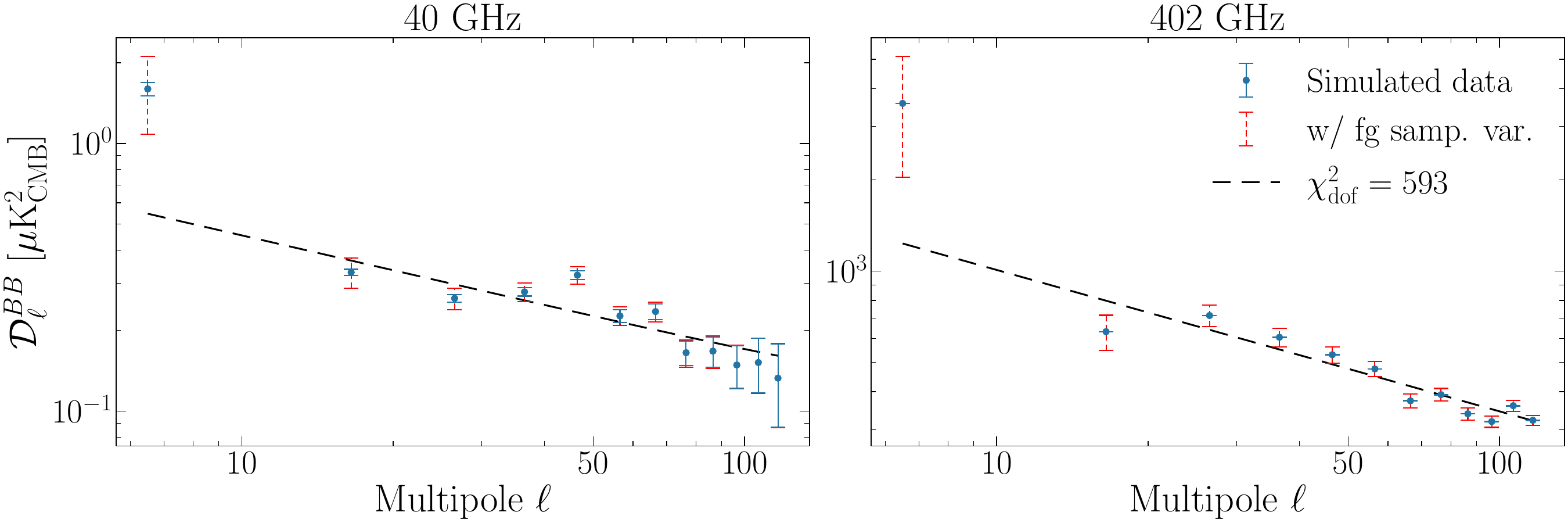}
    \caption{Mean $BB$ angular power spectra computed from the $N_{\rm sim}$ simulated maps at $40$ and $402$~GHz for the \texttt{d9s4} Galactic emission model. Blue error bars correspond to the covariance matrix computed in section~\ref{sec:covmat} after subtracting the foreground sample variance. Uncertainties obtained when accounting for this additional contribution are shown as dashed red error bars. Dashed lines indicate the best-fit power-law model for the synchrotron (left) and dust (right) power spectra assuming no foreground auto-correlations. The value $\chi^2_{\rm dof} = 593$ is obtained from the joint fits to the $N_{\rm cross}$ cross-frequency angular power spectra.}
    \label{fig:PL_spectra}
\end{figure}

We first model the $\ell$-dependence of the polarized dust and synchrotron emission angular power spectra using the power law of eq.~\eqref{eq:PL_spectra}. This model is regularly used for \textit{SO}~\cite{Azzoni2021, Azzoni2023, Hensley2022}, given its lower sensitivity and smaller sky coverage compared to \textit{LiteBIRD}. The resulting fits to the $B$-mode spectra at $40$ and $402$~GHz for \texttt{d9s4} are shown in figure~\ref{fig:PL_spectra}. In this plot, we show in blue the errors derived from the covariance matrix computed in section~\ref{sec:covmat} after subtracting the foreground sample variance. This constitutes the baseline approach adopted in this paper and is used to derive all the results presented hereafter. Since there is ongoing debate as to whether foreground auto-correlations should be included when analyzing real data~\cite{Vacher2022a, Hensley2022, Abril-Cabezas2024}, we also show, for comparison, the corresponding uncertainties obtained when this additional contribution is taken into account, as dashed red error bars. Providing a definitive answer to this question is beyond the scope of the present work, and we aim to compare these two estimates and investigate their implications in future work. In both cases, the model clearly fails to reproduce the data at \textit{LiteBIRD} sensitivity. Indeed, after subtracting the foreground sample variance (blue error bars), the average chi-square per degree of freedom is $\chi^2_{\rm dof} = 593$, indicating that any departure from a power law leads to significantly degraded fits. This discrepancy is particularly strong at high frequencies and, at low frequencies, for $\ell \lesssim 50$, showing that neither polarized thermal dust nor synchrotron power spectra can be described by a power law at \textit{LiteBIRD} sensitivity. At $\nu = 402$~GHz and $\ell = 26.5$, this results in a mismatch of $\sim 70$~$\mu$K$_{\rm CMB}^2$ between the measured and recovered power spectra, which is six orders of magnitude above the expected CMB $B$-mode signal at that angular scale. \par

For this reason, ref.~\cite{Vacher2022a} did not adopt this parametrization to obtain an unbiased estimate of the tensor-to-scalar ratio from \textit{LiteBIRD} simulations. The same effect was already hinted at by \textit{Planck} on the largest scales ($\ell \lesssim 40$), depending on the analyzed sky region~\cite{Planck2018_XI}, and is also observed in \textit{SO} simulations~\cite{Liu2025}, highlighting the necessity for future CMB missions to include additional free parameters to capture the full complexity of polarized foregrounds. \par

In the remainder of this work, we therefore fit the dust and synchrotron power spectra using independent amplitudes in each multipole bin.

\subsection{Spectral complexity of polarized Galactic emission} \label{sec:model_comparison}

Fitting $\ell$-by-$\ell$ amplitudes for the thermal dust and synchrotron power spectra, rather than using the power-law model of eq.~\eqref{eq:PL_model} as discussed in section~\ref{sec:PL_spectra}, yields satisfactory fits to the simulated data. In this framework, we study the potential of \textit{LiteBIRD} to reveal spatial variations in physical properties of Galactic foregrounds, specifically by detecting variations in   the spectral parameters: the dust temperature $T_{\rm d}(\ell)$, dust and synchrotron spectral indices at different scales $\beta_{\rm d}(\ell)$ and $\beta_{\rm s}(\ell)$, amplitudes $A_{\ell, \rm d}^{XX}$ and $A_{\ell, \rm s}^{XX}$, as well as the correlation coefficient between dust and synchrotron radiation $\rho_\ell^{XX}$. To simplify the discussion, in the following we focus only on the results obtained with the $BB$ cross-frequency angular power spectra. The corresponding plots for $EE$ can be found in appendix~\ref{app:E_figures}, and the plots showing the fits to the $B$-mode power spectra in \textit{Planck} frequency channels can be found in appendix~\ref{app:planck_figures}. \par

Figure~\ref{fig:models} shows the means and standard deviations of the fitted parameters in the $B$-mode cross-frequency angular power spectra, obtained in each multipole bin for the three Galactic emission models discussed in this article, \texttt{d9s4}, \texttt{d10s5} and \texttt{d12s7}. The fitted dust and synchrotron power spectra amplitudes $A_{\ell, \rm d}^{BB}$ and $A_{\ell, \rm s}^{BB}$ at $\nu_{\rm d} = 402$~GHz and $\nu_{\rm s} = 40$~GHz, respectively, are plotted in the upper panels and are consistent with the values that can be computed from the template maps from which the models are derived (in dashed lines). The middle left panel clearly demonstrates the ability of \textit{LiteBIRD} to constrain key parameters such as $\rho_\ell$. Indeed, for all scenarios, we retrieve the input values with error bars as low as $\sigma(\rho) \sim 10^{-2}$ for the largest angular scales. Polarized dust and synchrotron emission were found by \textit{Planck} to be highly correlated on large scales~\cite{Planck2018_XI}. However, \textit{Planck} measurements of synchrotron emission were dominated by noise for $\ell \gtrsim 50$, making it challenging to determine its correlation with polarized dust. The same effect is reproduced in our simulations of \textit{Planck} frequency bands, as it is shown in figure~\ref{fig:models_planck}. Figure~\ref{fig:models} shows that \textit{LiteBIRD} will help extend these measurements to smaller angular scales.

\begin{figure}[t]
    \centering
    \includegraphics[width=\linewidth]{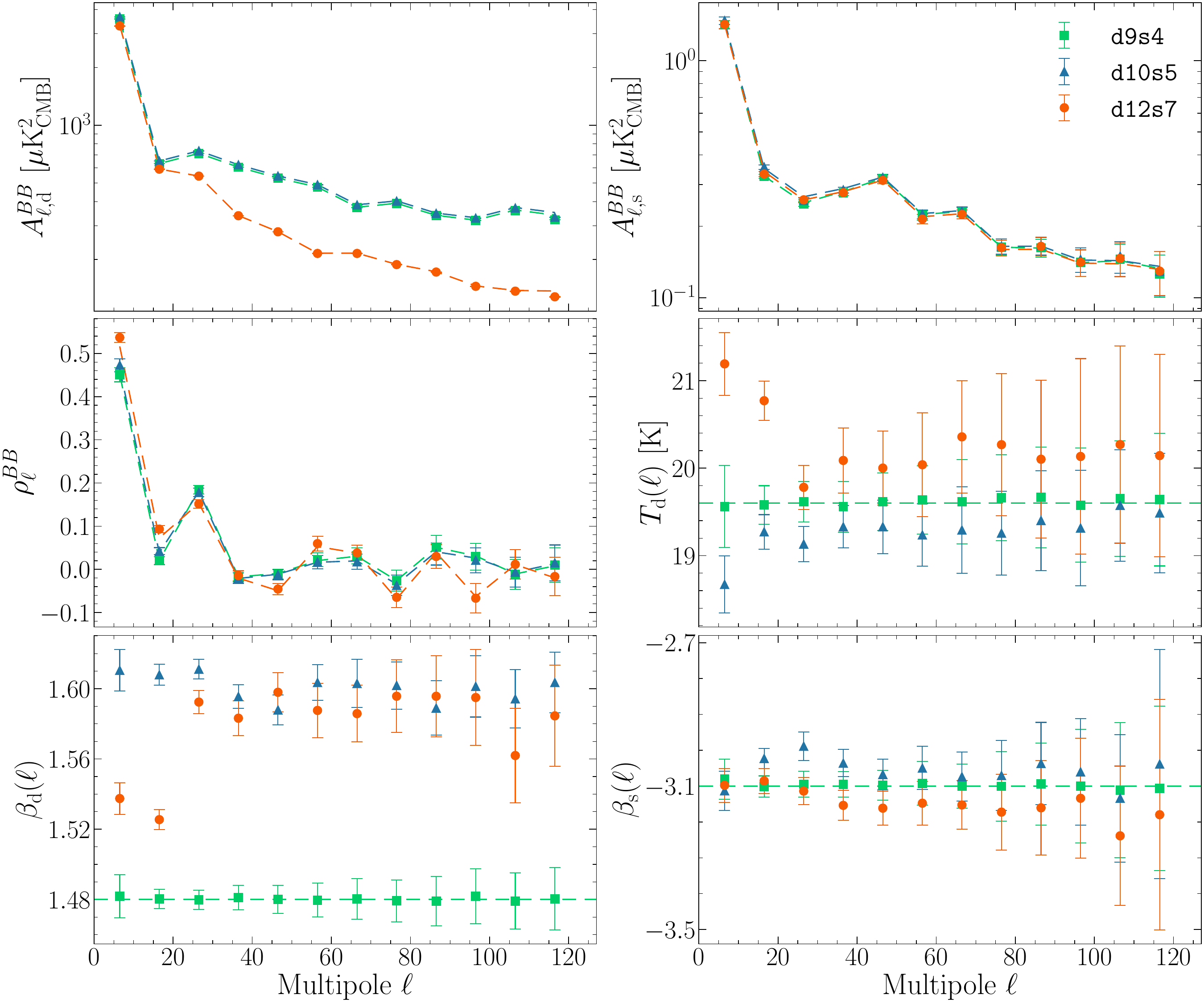}
    \caption{Dust and synchrotron amplitudes, correlation coefficient and spectral parameters inferred from $\ell = 2$ to $\ell = 121$ for $f_{\rm sky} = 0.7$ with \textit{LiteBIRD} instrumental characteristics, for the \texttt{d9s4} (green squares), \texttt{d10s5} (blue triangles) and \texttt{d12s7} (orange circles) Galactic emission models in the case of the $B$-mode power spectra. The colored dashed lines represent the input parameters for each scenario, when it is defined.}
    \label{fig:models}
\end{figure}

As expected, in the case of the \texttt{d9s4} model, we find constant values for the dust temperature, as well as for the dust and synchrotron spectral indices, over the whole multipole range. We recover as desired the original values of the low complexity model in each bin of $\ell$, namely $T_{\rm d}(\ell) = 19.6$~K, $\beta_{\rm d}(\ell) = 1.48$ and $\beta_{\rm s}(\ell) = -3.1$ (see section~\ref{sec:fg}). For more complex models, this is no longer the case. As spectral parameters vary across the sky, the fitted parameters vary across multipoles, and even more strongly in the case of \texttt{d12s7}, where dust spectral parameters also vary along the LOS. \par

\begin{figure}[t]
    \centering
    \includegraphics[width=0.5\linewidth]{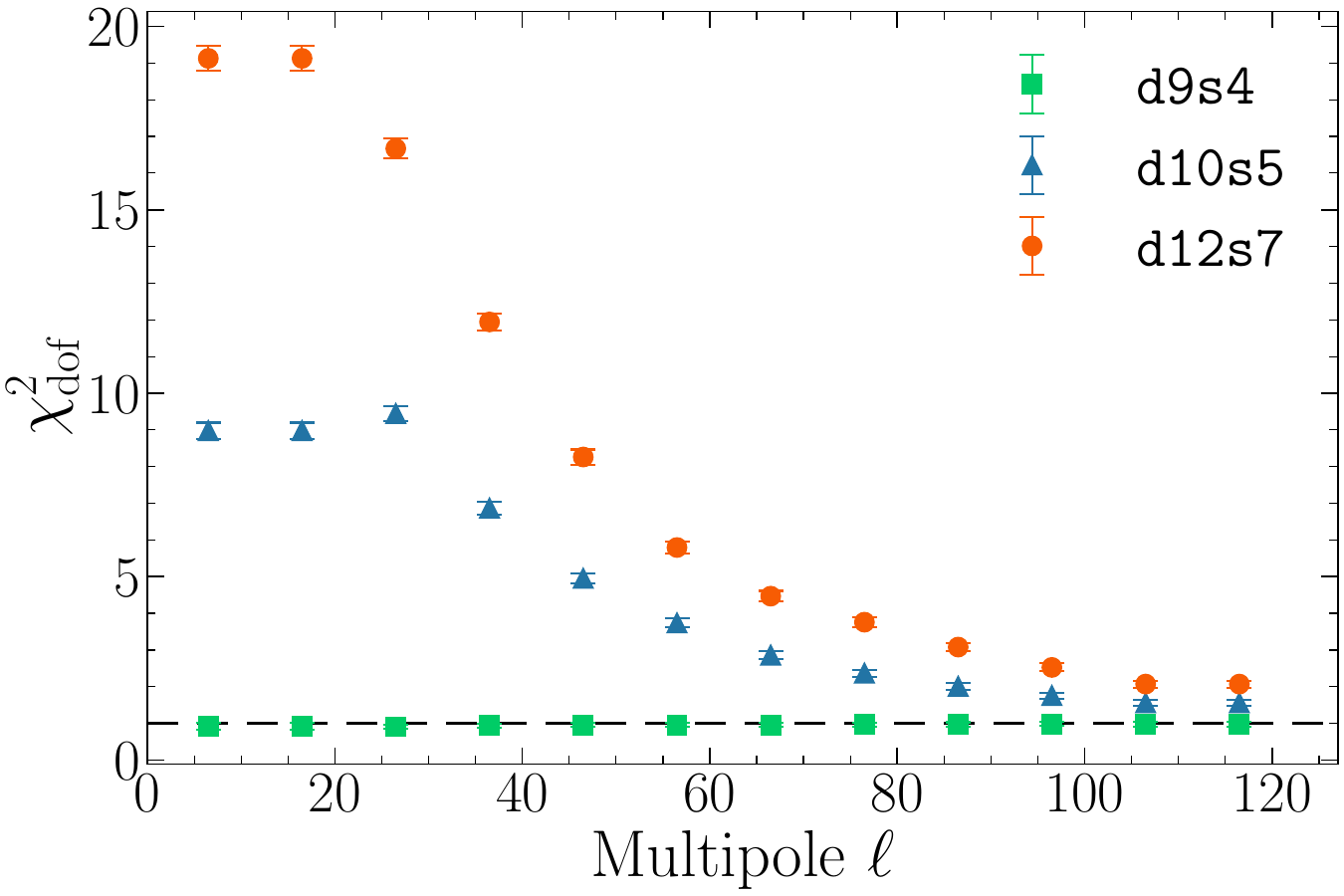}
    \caption{Chi-square per degree of freedom obtained from the fits of the \texttt{d9s4} (green squares), \texttt{d10s5}~(blue triangles) and \texttt{d12s7} (orange circles) Galactic emission models in each multipole bin for the $B$-mode power spectra. As they are fit together (see section~\ref{sec:fits}), the first and last bin presents the same $\chi^2_{\rm dof}$ as the second and penultimate one, respectively. The dashed line shows the reference value of $\chi^2_{\rm dof} = 1$.}
    \label{fig:chi2r}
\end{figure}

These variations are accompanied by a strong increase of the chi-square per degree of freedom, $\chi^2_{\rm dof}$, for the corresponding fits, see figures~\ref{fig:chi2r} and \ref{fig:chi2r_EE} for the $B$-mode and the $E$-mode power spectra, respectively. While the fits of the \texttt{d9s4} Galactic emission model are compatible with $\chi^2_{\rm dof} = 1$ at all scales, the reduced chi-square at $\ell = 6.5$ is increased to $\chi^2_{\rm dof} \sim 9$~($\sim 6$ for $EE$) in the \texttt{d10s5} model and $\chi^2_{\rm dof} \sim 19$~($\sim 11$ for $EE$) in the case of \texttt{d12s7}, showing the incompatibility of the simple parametric model of eq.~\eqref{eq:model} with the simulated \textit{LiteBIRD} data containing higher spectral complexity Galactic emission (the latter remaining sufficient at \textit{Planck} sensitivities, see figure~\ref{fig:chi2r_planck}). This incompatibility arises from the SED distortions introduced by models containing spectral parameters that are not constant across the Galaxy. In these cases, the MBB and PL models for thermal dust and synchrotron emission are not a good description of the simulated data, which also results in enlarged error bars on the spectral parameters, as shown in figures~\ref{fig:models} and \ref{fig:models_EE}. The uncertainties on $A_{\ell, \rm d}^{XX}$, $A_{\ell, \rm s}^{XX}$ and $\rho_\ell^{XX}$, on the other hand, are much less affected, since these parameters mainly trace the amplitudes of the dust and synchrotron power spectra rather than the frequency scaling itself. This demonstrates the ability of \textit{LiteBIRD} to detect and quantify deviations from the aforementioned parametric model, arising from variations of the spectral parameters in the three dimensions of our Galaxy. To model these distortions, eq.~\eqref{eq:model} can be extended using a moment expansion of the SED, which will be discussed in section~\ref{sec:mom}. \par

As can be seen in the lower right panels of figure~\ref{fig:models}, deviations are detected in the value of $\beta_{\rm s}(\ell)$ between the \texttt{d10s5} Galactic emission model, containing no curvature in the synchrotron SED, and \texttt{d12s7} where the synchrotron SED is described by eq.~\eqref{eq:PL_curved}. As synchrotron radiation in these two Galactic emission models is generated from the same amplitude and spectral index maps, the observed differences could only be explained by the presence of $c_{\rm s}$ in the \texttt{s7} model, or by the dust increased spectral complexity in the \texttt{d12} model with respect to \texttt{d10}. By comparing simulations of \texttt{d9s5} and \texttt{d9s7}, we could verify that these observed variations can be entirely explained by the presence of curvature in the synchrotron SED.\footnote{Discrepancies between the best-fit values for $\beta_{\rm s}(\ell)$ are still observed at large angular scales between \texttt{d9s5} and \texttt{d9s7}, demonstrating that the dust increased complexity is not responsible for the observed effects.} However, \textit{LiteBIRD} covers frequencies that are not low enough to detect $c_{\rm s} \neq 0$ in the $EE$ and $BB$ cross-frequency angular power spectra, leading to poor fits due to the strong degeneracy between $\beta_{\rm s}(\ell)$ and $c_{\rm s}(\ell)$. These results highlight the fact that even if no detection is possible using \textit{LiteBIRD} data only, the potential presence of curvature in the synchrotron SED can have some effects on the recovery of $\beta_s(\ell)$. The impact of such an effect---which might be critical for component separation---could be mitigated by the use of complementary low-frequency data, for instance from \textit{C-BASS}~\cite{Harper2022} and \textit{QUIJOTE}~\cite{Rubino-Martin2023}, as well as priors on the spectral parameters~\cite{Jew2019}. \par

\begin{figure}[t]
    \centering
    \includegraphics[width=\linewidth]{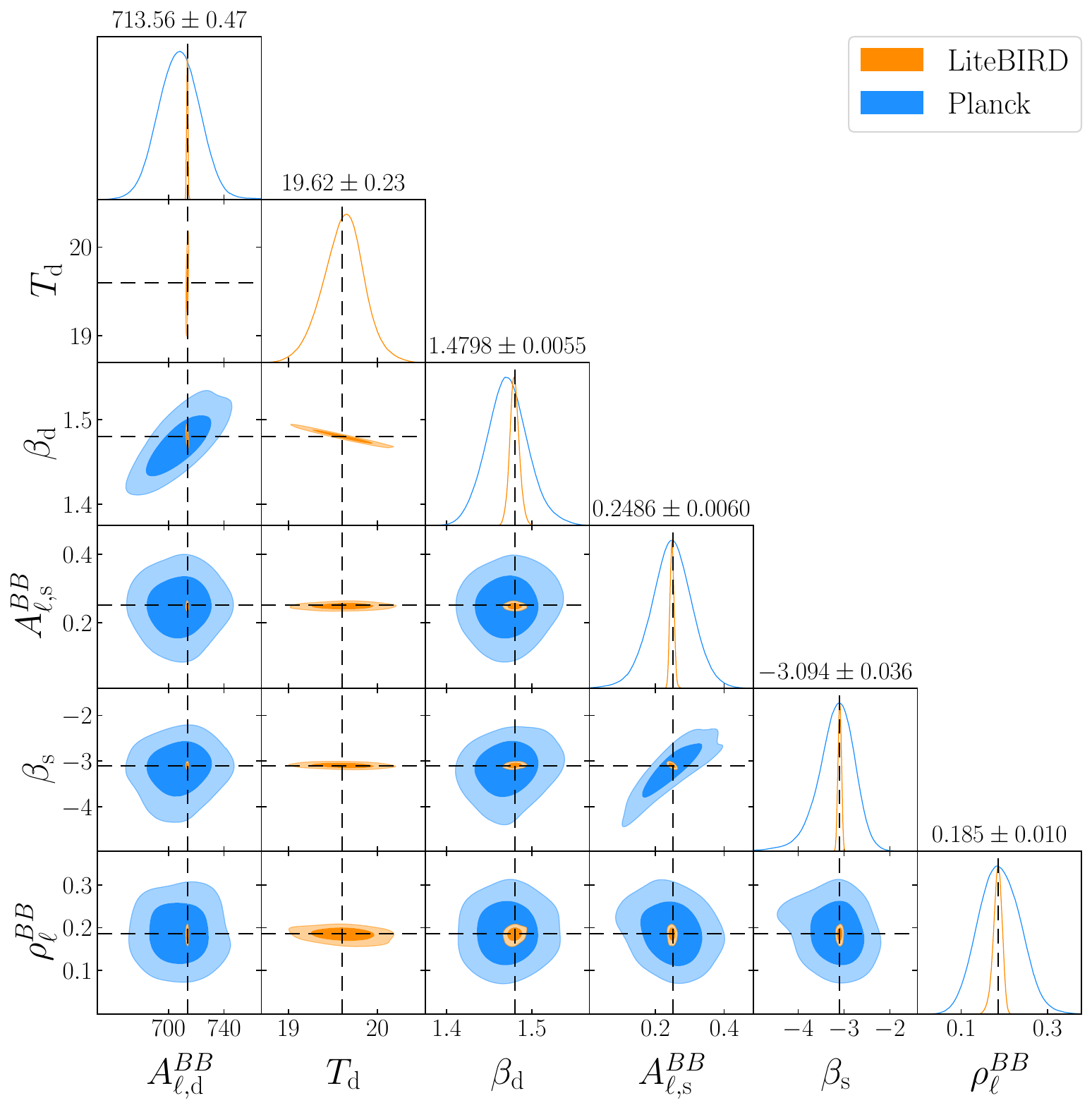}
    \caption{Sampling distribution of the best-fit parameter estimates obtained from $N_{\rm sim}$ simulations of $B$-mode cross-frequency angular spectra with the \texttt{d9s4} Galactic emission model, for the multipole bin covering $\ell = 22 \rightarrow 31$. Dark and light shaded contours enclose $68 \, \%$ and $95 \, \%$ of the recovered best-fit parameter estimates, respectively, for \textit{LiteBIRD} (orange) and \textit{Planck} (blue). The black dashed lines indicate the input values, and the inferred parameters for \textit{LiteBIRD} are quoted at the top of each column as $68 \, \%$ confidence intervals. The dust temperature $T_{\rm d}$ is fixed to its input value when fitting the power spectra from \textit{Planck} simulations.}
    \label{fig:PlvsLB}
\end{figure}

In order to assess the accuracy of \textit{LiteBIRD} measurements of the dust and synchrotron spectral parameters and amplitudes, we compute their distribution using \texttt{GetDist} in each multipole bin. Figure~\ref{fig:PlvsLB} shows the obtained distribution in the bin covering $\ell = 22 \rightarrow 31$ centered on $\ell = 26.5$, for the lowest complexity Galactic emission model. For comparison, we also compute the distribution obtained with \textit{Planck} instrumental characteristics in polarization at frequencies $30$, $44$, $70$, $100$, $143$, $217$ and $353$~GHz. We verified that the obtained error bars are consistent with \textit{Planck} 2018 results~\cite{Planck2018_XI}, with the differences that they used bandpowers of width $\Delta\ell=20$ instead of $\Delta\ell=10$ in this work, leading to factors $\sim \sqrt{2}$ between the forecast errors, and that we do not simulate complementary data from \textit{WMAP}~\cite{Bennett2013} to constrain low frequency foregrounds, leading to worse constraints on the synchrotron component in our case. In the multipole range considered in this analysis, these ancillary data lead to uncertainties of $\sigma(\beta_{\rm d}) = 0.03 \text{--} 0.05$, $\sigma(\beta_{\rm s}) = 0.05 \text{--} 0.18$, and $\sigma(\rho) = 0.01 \text{--} 0.03$ in $B$~modes, while the lower-frequency data from \textit{S-PASS} enable constraints on the synchrotron spectral index at the level of $\sigma(\beta_{\rm s}) \sim 0.1$~\cite{Krachmalnicoff2018}. \par

Figure~\ref{fig:PlvsLB} and \ref{fig:PlvsLB_EE} (see appendix~\ref{app:E_figures}) clearly show the improvement that will be brought by \textit{LiteBIRD} with respect to \textit{Planck}. Constraints on the dust and synchrotron $B$-mode ($E$-mode) amplitude are tightened by factors $32$ and $10$ ($33$ and $8.9$), respectively. This yields error bars of the correlation coefficient between thermal dust and synchrotron radiation to be tightened with respect to constraints obtained with \textit{Planck} sensitivities by a factor $4.8$~($3.3$). The dust and synchrotron spectral indices benefit from improvements by factors $4.4$ and $12$ ($2.0$ and $7.7$), yielding sensitivities per multipole bin of $\sigma(\beta_{\rm d}) \sim 0.006$ ($\sim 0.010$) and $\sigma(\beta_{\rm s}) \sim 0.04$ ($\sim 0.03$). Moreover, during the fits with \textit{Planck} sensitivities, we fix the value of the dust temperature to its input value, $T_{\rm d}(\ell) = 19.6$~K. Indeed, this parameter remains unconstrained if we let it free during the fitting process, similar to what is forecast for \textit{SO} in ref.~\cite{Hensley2022}. With \textit{LiteBIRD}, we anticipate obtaining tight constraints on the dust temperature, with uncertainties as low as $\sigma(T_{\rm d}) \sim 0.2$~K ($\sim 0.4$~K) at the largest scales. This stems from a combination of a higher maximum-frequency channel at $402$~GHz (instead of $353$~GHz for \textit{Planck}), better sensitivity, and improved spectral resolution.

\subsection[Searching for $E$/$B$- and $T$/$P$-discrepancies]{\boldmath Searching for $E$/$B$- and $T$/$P$-discrepancies} \label{sec:discrepancies}

We then focus on the potential of \textit{LiteBIRD} to find differences between $E$, $B$ and $T$  emission properties in complex Galactic emission models. Detecting such differences is of prime importance as it could be the signature of the existence of different populations of grains emitting in intensity and polarization, or the direct observable consequence of the co-variation of the GMF and the physical conditions in the three dimensions of our Galaxy. \par

\begin{figure}[t]
    \centering
    \includegraphics[width=\linewidth]{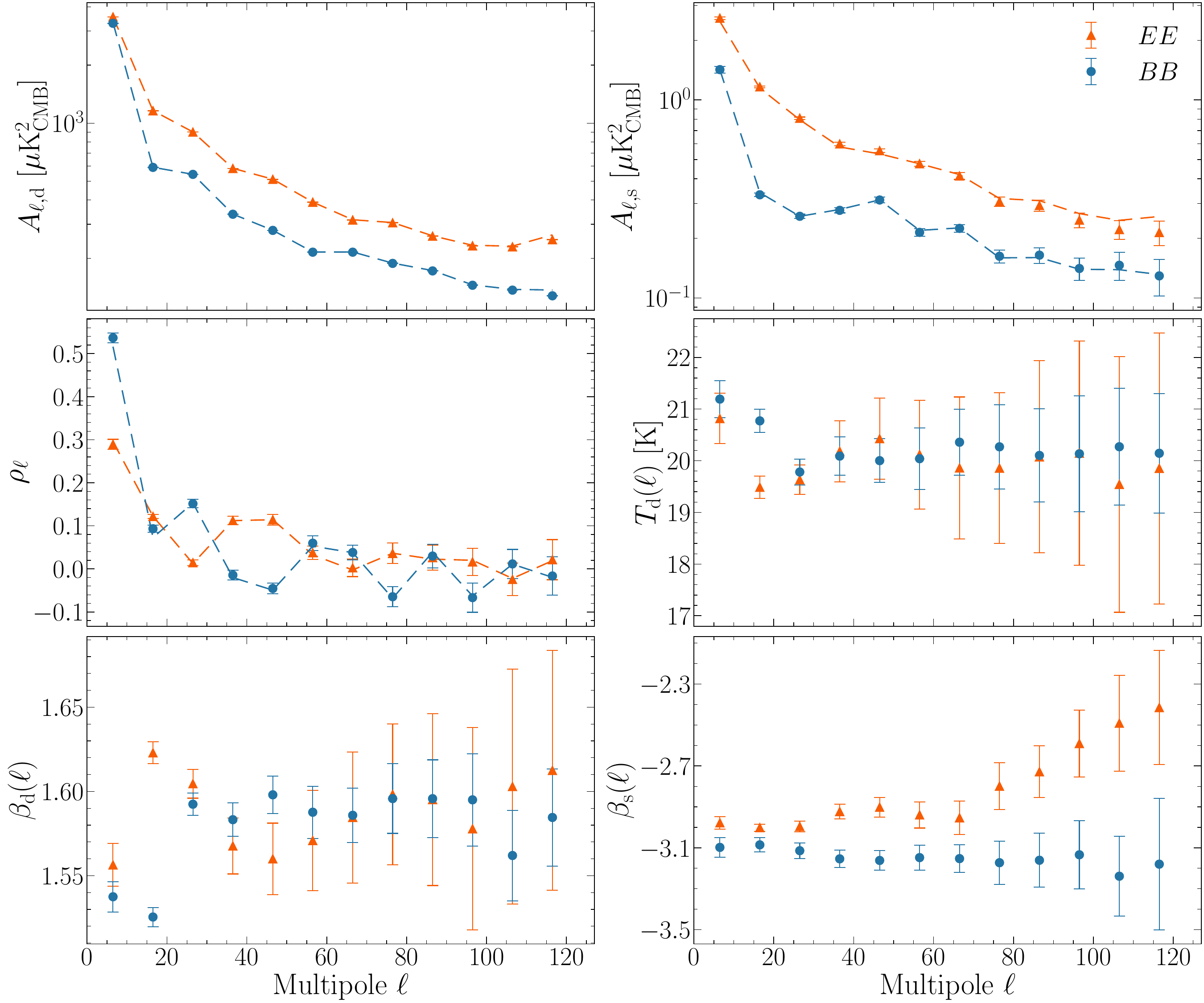}
    \caption{Dust and synchrotron amplitudes, correlation coefficient and spectral parameters inferred from the fits of the $E$-mode (orange triangles) and $B$-mode (blue circles) cross-frequency angular power spectra from $\ell = 2$ to $\ell = 121$ with \textit{LiteBIRD} instrumental characteristics, for the \texttt{d12s7} Galactic emission model with $f_{\rm sky} = 0.7$. The input parameters are shown by the colored dashed lines, when defined.}
    \label{fig:modes}
\end{figure}

Figure~\ref{fig:modes} displays the best-fit parameters and their uncertainties, obtained from the means and standard deviations of the fits to the $N_{\rm sim}$ simulations of $E$- and $B$-mode cross-frequency angular power spectra in each multipole bin for the high complexity model \texttt{d12s7}. In the middle left panel, one can see that thermal dust and synchrotron emission are more strongly correlated in $B$ modes than in $E$ modes for $\ell \leq 11$, consistent with what is observed in \textit{Planck} data~\cite{Planck2018_XI}. One can also notice that the dust temperature and spectral index are generally better constrained in $B$ modes than in $E$ modes, especially at large $\ell$. This arises from the full structure of the cross-frequency spectra covariance matrix itself, resulting from the combined effect of CMB cosmic variance, instrumental noise in each frequency band, and chance correlations with foregrounds. Indeed, from the diagonal of the $E$- and $B$-mode covariance matrices computed in section~\ref{sec:covmat}, we find the dust signal-to-noise ratio at $402$~GHz to be higher for $B$ than for $E$. Constraints on $\beta_{\rm s}$ are, however, similar for $E$ and $B$ modes, which we also understand from the $EE$ and $BB$ signal-to-noise ratios of synchrotron emission at $40$~GHz. From \textit{Planck} data (see appendix~\href{https://www.aanda.org/articles/aa/pdf/2020/09/aa32618-18.pdf\#page=29}{C} of ref.~\cite{Planck2018_XI}), constraints on $\beta_{\rm s}$ are better by a factor $\sim 2$ in $E$ modes than in $B$ modes whereas constraints on $\beta_{\rm d}$ are similar, in consistency with our simulated data (see figure~\ref{fig:modes_planck}). Beyond the increased sensitivity one can expect from \textit{LiteBIRD} in $E$-mode polarization with respect to other experiments, combining them with $B$ modes seems therefore to be a promising tool for Galactic science studies. \par

\begin{figure}[t]
    \centering
    \includegraphics[width=0.5\linewidth]{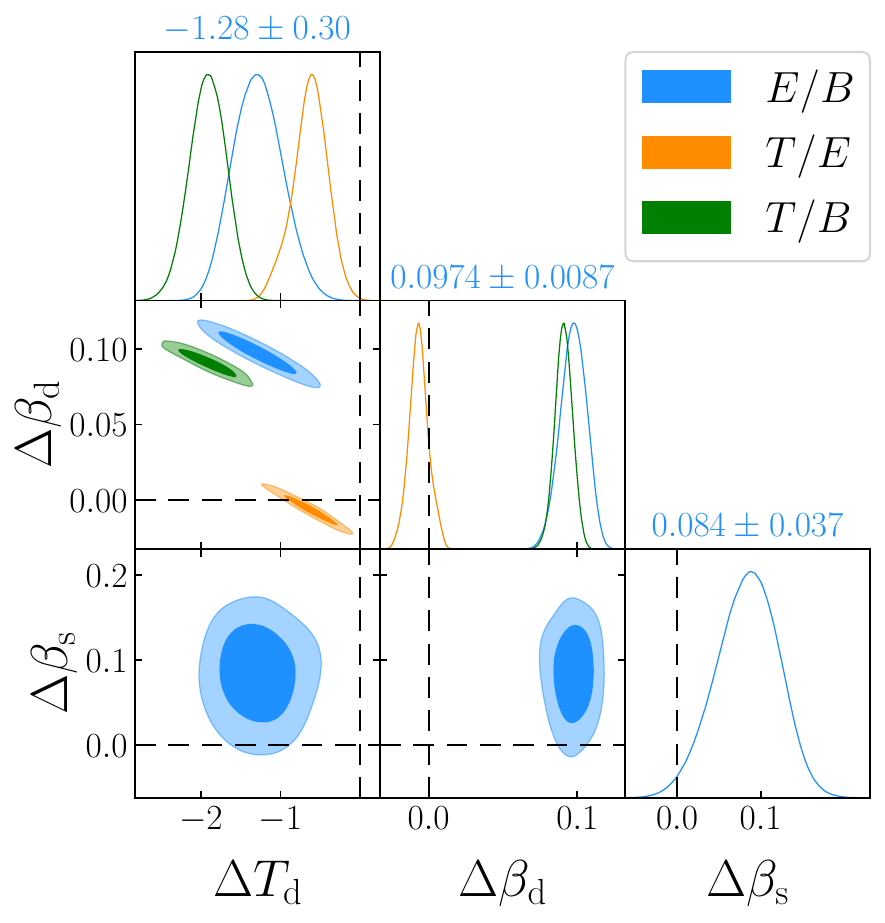}
    \caption{Differences between dust and synchrotron spectral parameters between $E$ and $B$ (blue), $T$ and $E$ (orange), and $T$ and $B$ (green) cross-frequency angular power spectra, as fitted from \textit{LiteBIRD} mock data containing the \texttt{d12s7} sky model at multipoles $\ell = 12 \rightarrow 21$. Dark and light areas represent the $68 \, \%$ and $95 \, \%$ confidence regions, respectively. $T$/$P$-discrepancies involving the synchrotron spectral index are not represented, since this parameter remains unconstrained at \textit{LiteBIRD} sensitivity in intensity. The quoted values on the diagonal correspond to the peaks and $68 \, \%$ confidence intervals of the marginalized distributions for $E$/$B$. The black dashed lines indicate the case with no discrepancy.}
    \label{fig:discrepancies}
\end{figure}

From figure~\ref{fig:modes}, one can see that differences between $E$ and $B$ modes tend to be observed at the largest scales for the spectral parameters $T_{\rm d}(\ell)$, $\beta_{\rm d}(\ell)$ and $\beta_{\rm s}(\ell)$, particularly in the second multipole bin. This $E$/$B$-discrepancy originates from the polarized mixing introduced by the variations of the dust and synchrotron physical properties across the sky, and can be modeled using the moment expansion formalism, as we will show in section~\ref{sec:mom}~\cite{Vacher2023}. This effect is visible in both \texttt{d10s5} and \texttt{d12s7} models, and is particularly important at $\ell = 16.5$. The sampling distribution of $T_{\rm d}^{EE} - T_{\rm d}^{BB}$, $\beta_{\rm d}^{EE} - \beta_{\rm d}^{BB}$ and $\beta_{\rm s}^{EE} - \beta_{\rm s}^{BB}$ in this bandpower is shown for \texttt{d12s7} in figure~\ref{fig:discrepancies}. In this multipole bin, we detect a $4.2 \, \sigma$ discrepancy between $E$ and $B$ for the dust temperature, a $11\, \sigma$ discrepancy for the dust spectral index, and the synchrotron spectral indices are found to be incompatible at the $2.2 \, \sigma$ level. We deduce that \textit{LiteBIRD} will be able to confidently detect $E$/$B$-discrepancy, highlighting variations of the physical conditions across the Galaxy. As shown in figure~\ref{fig:modes_planck}, discrepancies between the dust SED in $E$ and $B$ modes are not significantly detected in \textit{Planck} simulations, and could only be observed in data when including large portions of the Galactic plane in the analysis, where the emission is much more intense and the mixing more important~\cite{Ritacco2023}. \textit{LiteBIRD} will therefore enable us to extend such analyses to high Galactic latitudes. \par

The thermal dust SED spectral index was measured by \textit{Planck} to be marginally different in intensity and polarization~\cite{Planck2015_XXII, Planck2018_XI}. In addition to these intrinsic differences~\cite{Ysard2024}, polarized mixing is also expected to introduce discrepancies in the spectral parameters, with $TT$ differing from both $EE$ and $BB$, in the same way as between $E$ and $B$ modes (see section~\ref{sec:mom}). As \textit{LiteBIRD} will also be able to provide data in intensity, a question that naturally arises is therefore whether it will be possible to detect a $T$/$P$-discrepancy. To give insights into this question, we perform fits of the $TT$ cross-frequency angular power spectra computed from our simulations containing the \texttt{d12s7} Galactic emission model. At \textit{LiteBIRD} lowest frequency channels, one can expect free-free emission to be a significant source of radiation~\cite{Planck2018_I}. However, the latter traces the hot ISM with high electron density. We expect it to be subdominant at the high Galactic latitudes considered in this study and therefore exclude it from both the simulations and the fits~\cite{Dickinson2003}. We adopt a fixed value of the synchrotron spectral index $\beta_{\rm s} = -3.1$, which is poorly constrained in intensity using \textit{LiteBIRD} frequency channels. In figure~\ref{fig:discrepancies}, we compare the obtained results with the ones previously derived for polarization for the multipole covering $\ell = 11 \rightarrow 21$. We detect discrepancies between $TT$ and $BB$ power spectra at the $8.3 \, \sigma$ level for $T_{\rm d}$ and at the $15 \, \sigma$ level for $\beta_{\rm d}$. Between the $EE$ and $TT$ power spectra, these discrepancies decrease to $2.8 \, \sigma$ for $T_{\rm d}$ and $0.98 \, \sigma$ for $\beta_{\rm d}$. We argue that, since the input spectral parameters maps are identical for $I$, $Q$, and $U$, the observed discrepancies can only arise from spectral mixing, which acts differently in intensity and polarization. These results, which are similar for the \texttt{d10s5} model, tend to show that dust $T$/$P$-discrepancy originating from spectral mixing between different emitting regions will be detectable by \textit{LiteBIRD}. Moreover, even if it is not the case in the models studied above, such an effect can also arise from the presence of multiple grain populations with different shapes and chemical compositions, resulting in a frequency-dependent polarization fraction~\cite{Hensley2023, Ysard2024}. Studying the possibility of discriminating these intrinsic variations from those induced by spectral mixing is left for future work. \par

Ultimately, we recall that the likelihood of eq.~\eqref{eq:likelihood} used in this analysis is Gaussian. Although it is known to be a good approximation at intermediate and small scales, it does not provide an accurate description of the lowest multipoles. More appropriate approximations exist to account for these effects~\cite{HL2008, Mangilli2015, Galloni2025}, and we leave the study of their impact on cross-frequency angular power spectrum analyses for future work.


\subsection{Moment expansion modeling of Galactic emission} \label{sec:mom}

A powerful way to model the complexity of the SED of the cross-frequency angular power spectra, as well as the $E$/$B$- and $T$/$P$-discrepancies, is given by the moment expansion formalism, that allows to describe the SED distortions due to polarized mixing. This formalism was introduced in intensity by ref.~\cite{Chluba2017} and generalized to polarization by ref.~\cite{Vacher2022b}. It was extended and applied at the angular power spectrum level for \textit{Planck} $TT$ data~\cite{Mangilli2021}, as well as \textit{SO}~\cite{Azzoni2021, Liu2025} and \textit{LiteBIRD}~\cite{Vacher2022a, Fuskeland2023} in the context of parametric component separation for $B$ modes. In addition, moments play a key role in order to tackle foregrounds in several internal linear combination (ILC) methods~\cite{Remazeilles2021, Adak2021, Carones2024}. Moment expansion was further used to explain and model $E$/$B$-discrepancy~\cite{Vacher2023}, and proven to be a powerful tool for Galactic science in order to explain the variance of polarized dust spectral properties~\cite{Guillet2025}. \par

To test the benefit from extending our model with moment expansion, we replace the foreground SED of eq.~\eqref{eq:model} by eq.~(\href{https://www.aanda.org/articles/aa/pdf/2022/04/aa42664-21.pdf\#page=3}{8}) of ref.~\cite{Vacher2022a}, including all auto- and cross-spectra of dust and synchrotron moments up to first order. Such a model is fitted independently to $TT$, $EE$ and $BB$ cross-frequency angular power spectra, leading to different moment values for the three datasets. These observed differences are entirely due to the polarized mixing induced by the spatial variations of foreground spectral properties in the complex GMF structure, and are ultimately understood as the source of the $E$/$B$- and $T$/$P$-discrepancies. Indeed, expanding the value of a given spectral parameter $\Pi \in \{\beta_{\rm d}, T_{\rm d}, \beta_{\rm s}\}$ around a chosen pivot value $\bar{\Pi}$, one can show that its $\ell$-dependence is given at first order by
\begin{equation}
    \Pi(\ell) \simeq \bar{\Pi} + \delta\Pi(\ell)
    = \bar{\Pi} + \mathcal{D}_\ell^{A \times \omega_1^\Pi} / \mathcal{D}_\ell^{A \times A},
    \label{eq:beta_ell}
\end{equation}

where $\mathcal{D}_\ell^{A \times A} = A_\ell$ is the zeroth order amplitude of the considered power spectra SED, and $\mathcal{D}_\ell^{A \times \omega_1^\Pi}$ is its first-order moment with respect to $\Pi$~\cite{Mangilli2021}.

Following the aforementioned reference, the fits are performed iteratively: the pivot values are first fixed to initial guesses (namely, $\beta_{\rm d} = 1.48$, $T_{\rm d} = 19.6$~K, and $\beta_{\rm s} = -3.1$), all moments are fitted, and the pivots are then updated using eq.~\eqref{eq:beta_ell}. The procedure is repeated until the pivot values change by less than $1 \, \%$ between successive iterations. For the sake of clarity, we present our results here only for the \texttt{d12s7} model, which has the strongest distortions and discrepancies. We however verify that adding the moments performs equally well on simpler models. \par

\begin{figure}[t]
    \centering
    \includegraphics[width=0.5\linewidth]{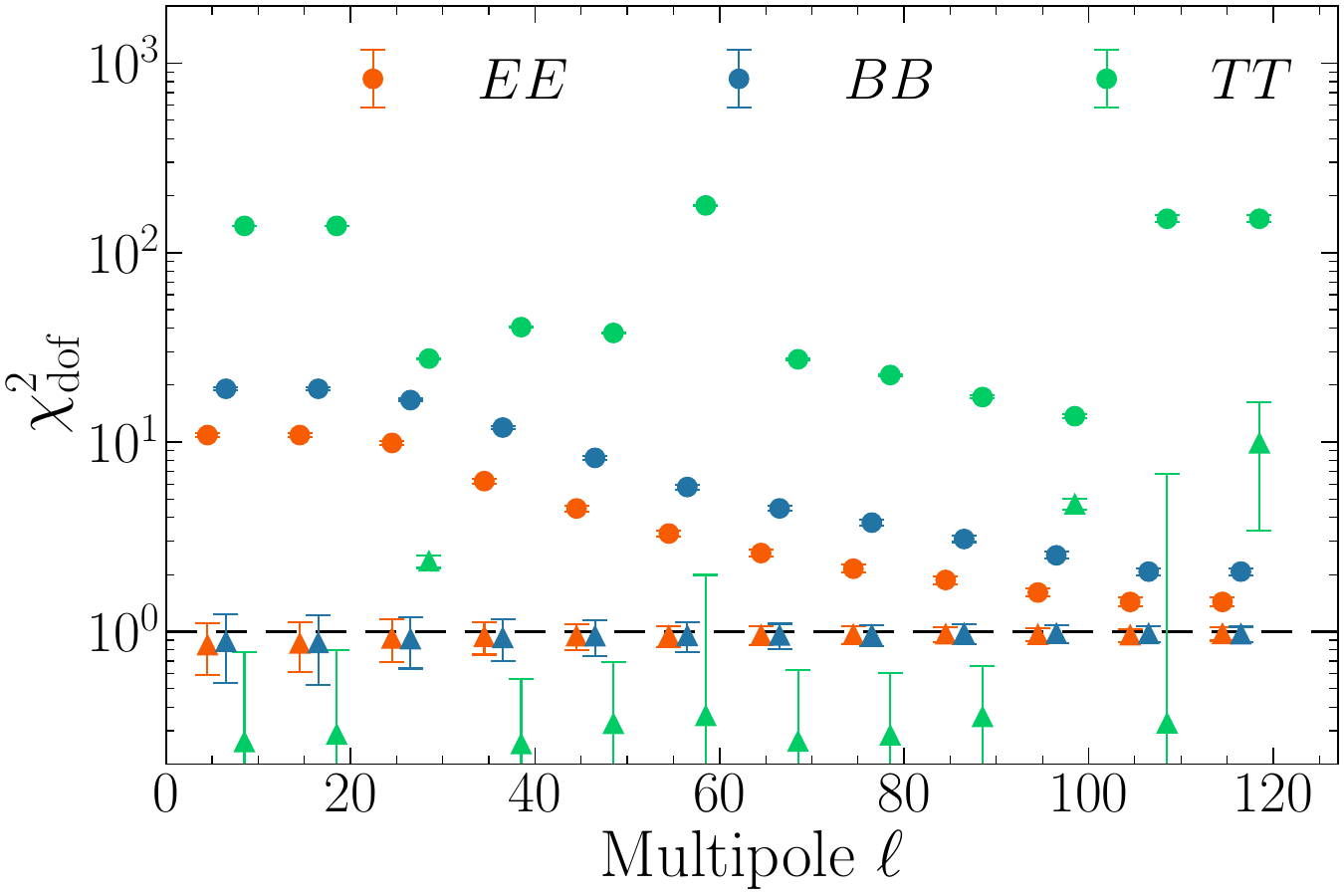}
    \caption{Reduced chi-square obtained from the fits of the \texttt{d12s7} simulations without (filled circles) and with (filled triangles) moment expansion, for $EE$ (orange), $BB$ (blue) and $TT$ (green) spectra. The reference value of $\chi^2_{\rm dof}=1$ is shown by the black dashed line.}
    \label{fig:chi2r_mom}
\end{figure}

In figure~\ref{fig:chi2r_mom}, we show a comparison between the reduced $\chi^2$ obtained using the model of eq.~\eqref{eq:model} (filled circles), and the one obtained with its extension with moment expansion (filled triangles). We notice that adding the moments renders the reduced $\chi^2$ compatible with~$1$ for each spectrum, thus providing a better modeling of the signal than the simple modified black-body plus power-law (MBB+PL) parametrization. The $TT$ covariance matrix is ill-conditioned due to strong correlations between spectra induced by CMB cosmic variance. A constant regularization term, whose amplitude is kept below $1 \, \%$ of the mean variance, is therefore added to each diagonal element before inversion to ensure numerical stability. This regularization artificially increases the covariance amplitude, resulting in values of $\chi^2_{\rm dof}$ below unity. \par

Using the methodology presented in ref.~\cite{Vacher2025}, one can infer, for a given pivot parameter, the theoretical values of the moment maps for each \texttt{PySM} model, from which the expected value of each moment coefficient at the angular power spectrum level can be derived. From this, one can compute theoretical values of the spectral parameters and their $T$/$P$- and $E$/$B$-discrepancies, using eq.~\eqref{eq:beta_ell} to estimate $\beta_{\rm d}^{XX}(\ell)$, $T_{\rm d}^{XX}(\ell)$ and $\beta_{\rm s}^{XX}(\ell)$ for each mode $XX \in \{TT, EE, BB\}$. These theoretical values can be compared to the results of the fits, giving insights into their ability to accurately recover the discrepancies emerging from the joint variations of spectral parameters and polarization angles in the Galaxy. \par

\begin{figure}[t]
    \centering
    \includegraphics[width=\linewidth]{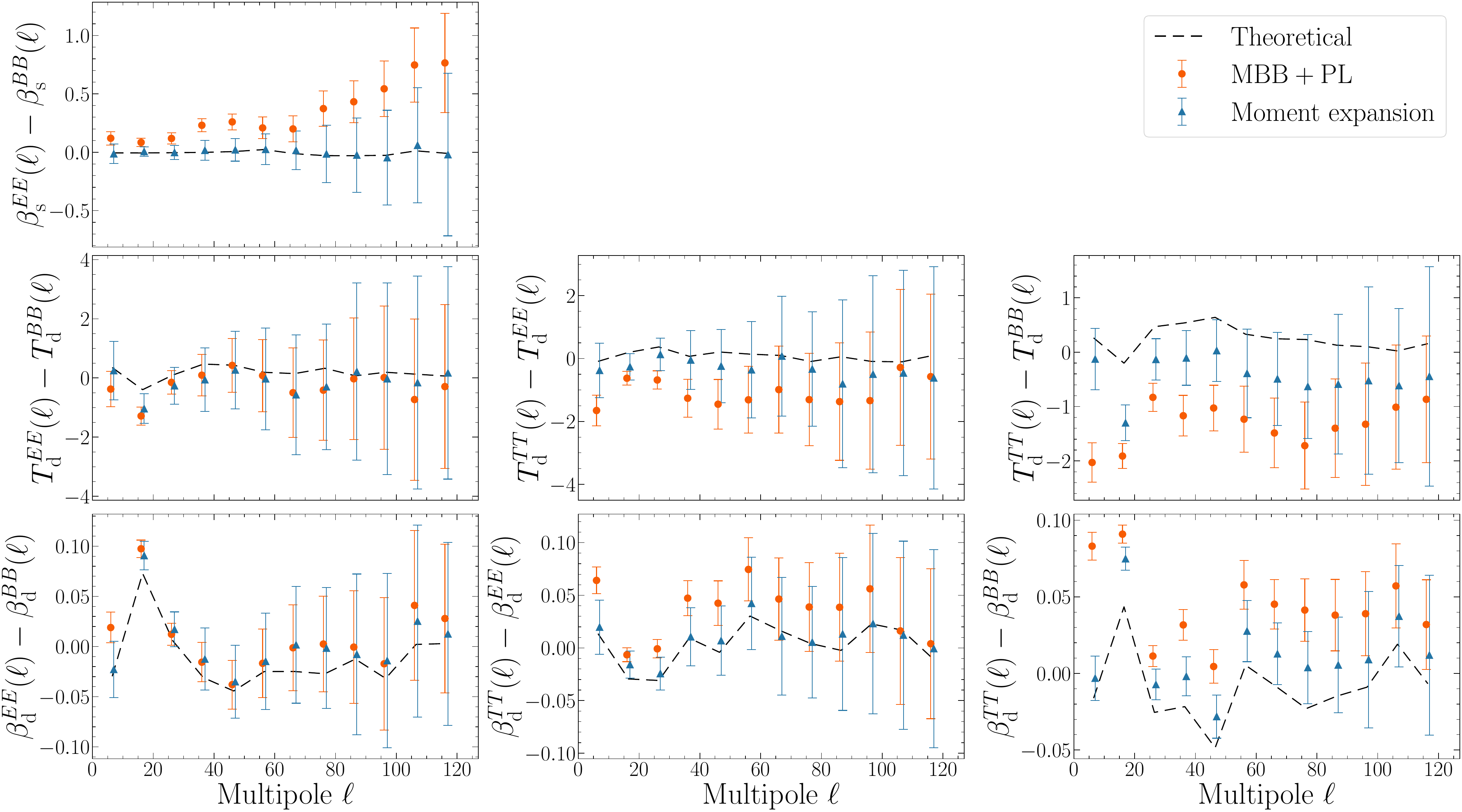}
    \caption{Inferred values of the $E$/$B$- (left column), $T$/$E$- (middle column) and $T$/$B$- (right column) discrepancies for the \texttt{d12s7} sky model. Dashed lines show the theoretical expectations obtained using the methodology presented in ref.~\cite{Vacher2025}. The fitted values using the zeroth order model are shown as orange circles, and the ones obtained from iterative fits of moments as blue triangles. The synchrotron spectral index is fixed to $\beta_{\rm s}=-3.1$ when fitting the $TT$ spectra.}
    \label{fig:discrepancies_moments}
\end{figure}

The results are displayed in figure~\ref{fig:discrepancies_moments}, both fitting the zeroth order MBB+PL model of eq.~\eqref{eq:model} (orange circles) and the iterative moment expansion extension (blue triangles). We clearly see that the theoretical discrepancies are recovered less precisely but more accurately with moment expansion than with the simpler model, especially in the cases where the $TT$ spectra, presenting higher signal-to-noise ratios, are involved,  as the presence of moments of the SED is even more significant in these cases. Indeed, moment expansion enables the recovery of the theoretical values at the $1 \, \sigma$ level for most of the multipole bins. Regarding the values of the spectral parameters themselves, adding the moments enables us to recover precisely their variations with multipole around their mean. However, the average fitted values are still systematically biased, this effect being particularly strong for the most complex foreground models (see figure~\ref{fig:beta_ell} for the dust $B$-mode spectral parameters in \texttt{d12s7}). Indeed, because of the strong distortions and correlations between parameters, moment expansion at first order does not completely succeed to model all the sky complexity. In appendix~\ref{app:high_order_moments}, we discuss the possibility of including second-order terms in the analysis to mitigate this effect. \par

Using the moment expansion formalism, \textit{LiteBIRD} will therefore be able to measure and detect the moment spectra beyond the simple SED parametrization, enabling quantification of the amplitude of the mixing and of the resulting discrepancies at different angular scales. However, we conclude by emphasizing that eq.~\eqref{eq:beta_ell} is only an approximation at first order. Indeed, any linear combination of MBB spectra cannot simply be parametrized as a resulting MBB with an effective $\Pi(\ell)$. The full description of such a combination is given only by the full moment expansion including all moments, resulting in frequency-dependent spectral parameters when eq.~\eqref{eq:beta_ell} is extended to higher orders. Therefore, even if the SED assumed for each dust (synchrotron) component is the canonical MBB (PL), future studies involving a mixture of these components may not be able to recover full, unbiased information without extending the model with moments.

\section{Discussion and conclusion} \label{sec:conclusion}

With its $15$ frequency bands ranging from $40$ to $402$~GHz, \textit{LiteBIRD} is a promising experiment to study the polarization of the ISM of the Milky Way. Thanks to its improved frequency coverage and sensitivities with respect to \textit{Planck} and other current experiments such as \textit{SO}, it will be able to accurately measure thermal dust and synchrotron radiation properties, giving new insights into the processes behind Galactic polarized emission. Despite the leap in our understanding of the magnetized ISM provided by \textit{Planck}, several critical open questions remain, questions to which \textit{LiteBIRD} might provide a definitive answer: how do the properties of dust and synchrotron emission depend on the structure of the GMF? What is the composition of interstellar dust? \par

In this work, we showed that \textit{LiteBIRD} will be able to accurately measure the polarized dust and synchrotron amplitudes from large to small scales, as well as the spatial correlation between these two processes with accuracies down to $\sigma(\rho) \sim 10^{-2}$. \textit{LiteBIRD} will also provide precise measurements of the dust and synchrotron spectral indices with error bars as low as $\sigma(\beta_{\rm d}) \sim 0.006$ and $\sigma(\beta_{\rm s}) \sim 0.04$ at the largest scales, as well as constraints on the dust temperature for the first time in polarization with microwave data, reaching $\sigma(T_{\rm d}) \sim 0.2$~K. \par

However, for complex foreground models, intensity and polarization cross-frequency angular power spectra at \textit{LiteBIRD} sensitivity are not compatible with the simple parametrization of eq.~\eqref{eq:model}, as was the case for past experiments. This arises from the averaging of SEDs with spatially varying spectral parameters along the line of sight and across the beam, as well as in harmonic space. This induces an $\ell$-dependence of the spectral indices and dust temperature, which is expected to differ between $TT$, $EE$, and $BB$ due to the co-variation of the emission properties and the magnetic field structure throughout the ISM. We showed that \textit{LiteBIRD} will be able to confidently detect these $E$/$B$- and $T$/$P$-discrepancies at high Galactic latitude, thanks to its wide frequency coverage and improved sensitivity relative to past experiments such as \textit{Planck}. \par

The natural step to go beyond eq.~\eqref{eq:model} is the moment expansion of the SED (see section~\ref{sec:mom}), aiming at modeling the underlying SED distortions using a Taylor expansion around the MBB parametrization. We showed that, with \textit{LiteBIRD} sensitivity, it becomes possible to detect the moment coefficients, which provide an accurate characterization of the Galactic signal. We also showed that moments provide the proper framework for interpreting and understanding the observed $E/B$- and $T/P$-discrepancies, as they allow us to predict the values recovered in the fits from the underlying properties of the foreground models---and, alternatively, to gain insights into the statistical properties of polarized mixing.

$T$/$P$-discrepancy can also originate from the mixing of several grain populations in diffuse dust clouds presenting different internal chemical compositions, size distributions, temperatures and alignment mechanisms, such that different spectral behaviors can be witnessed in intensity and polarisation. We expect that $LiteBIRD$, equipped with the moment expansion machinery, would be able to  study such a scenario, as long as dust populations emit as MBBs. However, a mismodeling of the SED, even including moments, could result from other effects than mixing, such as the intrinsic deviations from the simple SED models we assumed in each emitting region of the Galaxy. In the case of dust, intrinsic deviations from the MBB could arise from the complex physics of grains, leading to a more complex (and perhaps not even analytical) intrinsic SED~\cite{Demyk2022}. Other complementary frameworks, such as one based on the analysis of the frequency covariance matrices of the maps~\cite{Guillet2025}, could extend the present study and further help assess \textit{LiteBIRD}’s ability to recover the statistical properties underlying the Galactic SED. We are aiming to explore this still open question in future work. Combined with analyses of the thermal dust SED directly in the pixel space, this approach promises to bring a detailed understanding of the processes underlying Galactic polarized emission. For example, we anticipate studying the ability of \textit{LiteBIRD} to discriminate between state-of-the-art dust models, such as Astrodust+PAH~\cite{Hensley2023} and THEMIS~2.0~\cite{Ysard2024}, in a future publication. \par

Often considered as a contamination for astronomical observations, attenuating and reddening starlight as well as masking extragalactic sources and the CMB, interstellar dust is full of complex physics, and our understanding of its physical and chemical properties is only in its earliest infancy. Together with Galactic synchrotron emission, they trace the structure of the ISM itself, organizing into filaments coupled with the GMF. Future CMB experiments such as \textit{LiteBIRD} will use their unprecedented sensitivities to greatly improve our understanding of the polarized emission of our Galaxy, and these advances will be crucial for subtracting the Galactic foregrounds from the primordial $B$-mode signal in the quest for new fundamental physics in the outer confines of the Universe.

\acknowledgments
%
This work is supported in Japan by ISAS/JAXA for Pre-Phase A2 studies, by the acceleration program of JAXA research and development directorate, by the World Premier International Research Center Initiative (WPI) of MEXT, by the JSPS Core-to-Core Program of A. Advanced Research Networks, and by JSPS KAKENHI Grant Numbers JP15H05891, JP17H01115, and JP17H01125.
The Canadian contribution is supported by the Canadian Space Agency.
The French \textit{LiteBIRD} phase A contribution is supported by the Centre National d’Etudes Spatiale (CNES), by the Centre National de la Recherche Scientifique~(CNRS), and by the Commissariat à l’Energie Atomique (CEA).
The German participation in \textit{LiteBIRD} is supported in part by the Excellence Cluster ORIGINS, which is funded by the Deutsche Forschungsgemeinschaft (DFG, German Research Foundation) under Germany’s Excellence Strategy (Grant No. EXC-2094 - 390783311).
The Italian \textit{LiteBIRD} phase A contribution is supported by the Italian Space Agency (ASI Grants No. 2020-9-HH.0 and 2016-24-H.1-2018), the National Institute for Nuclear Physics (INFN) and the National Institute for Astrophysics (INAF).
Norwegian participation in \textit{LiteBIRD} is supported by the Research Council of Norway (Grant No. 263011 and 351037) and has received funding from the European Research Council (ERC) under the Horizon 2020 Research and Innovation Programme (Grant agreement No. 772253, 819478, and 101141621).
The Spanish \textit{LiteBIRD} phase A contribution is supported by MCIN/AEI/10.13039/501100011033, project refs. PID2019-110610RB-C21, PID2020-120514GB-I00, PID2022-139223OB-C21, PID2023-150398NB-I00 (funded also by European Union NextGenerationEU/PRTR), and by MCIN/CDTI ICTP20210008 (funded also by EU FEDER funds).
Funds that support contributions from Sweden come from the Swedish National Space Agency (SNSA/Rymdstyrelsen) and the Swedish Research Council (Reg. no. 2019-03959).
The UK  \textit{LiteBIRD} contribution is supported by the UK Space Agency under grant reference ST/Y006003/1 - "LiteBIRD UK: A major UK contribution to the LiteBIRD mission - Phase1 (March 25)."
The US contribution is supported by NASA grant no. 80NSSC18K0132.
%
SV would like to acknowledge Mathias Regnier for useful discussions during this work.
JARM acknowledges financial support from the Horizon Europe research and innovation program under GA 101135036 (RadioForegroundsPlus).
The authors acknowledge the use of the \texttt{healpy}~\cite{Gorski2005, Zonca2019}, \texttt{PySM}~\cite{Thorne2017, Zonca2021, Panexp2025}, \texttt{CAMB}~\cite{Lewis2000}, \texttt{NaMaster}~\cite{Alonso2019}, \texttt{mpfit}~\cite{Markwardt2009}, \texttt{GetDist}~\cite{Lewis2025}, \texttt{NumPy}~\cite{Harris2020}, \texttt{SciPy}~\cite{Virtanen2020}, \texttt{Astropy}~\cite{Astropy2013, Astropy2018, Astropy2022}, and \texttt{Matplotlib}~\cite{Hunter2007} software packages.




\bibliographystyle{JHEP}
\bibliography{biblio}

\appendix


\section{Effect of the sky fraction on the fits} \label{app:fsky_comparison}

\begin{figure}[t]
    \centering
    \includegraphics[width=0.49\linewidth]{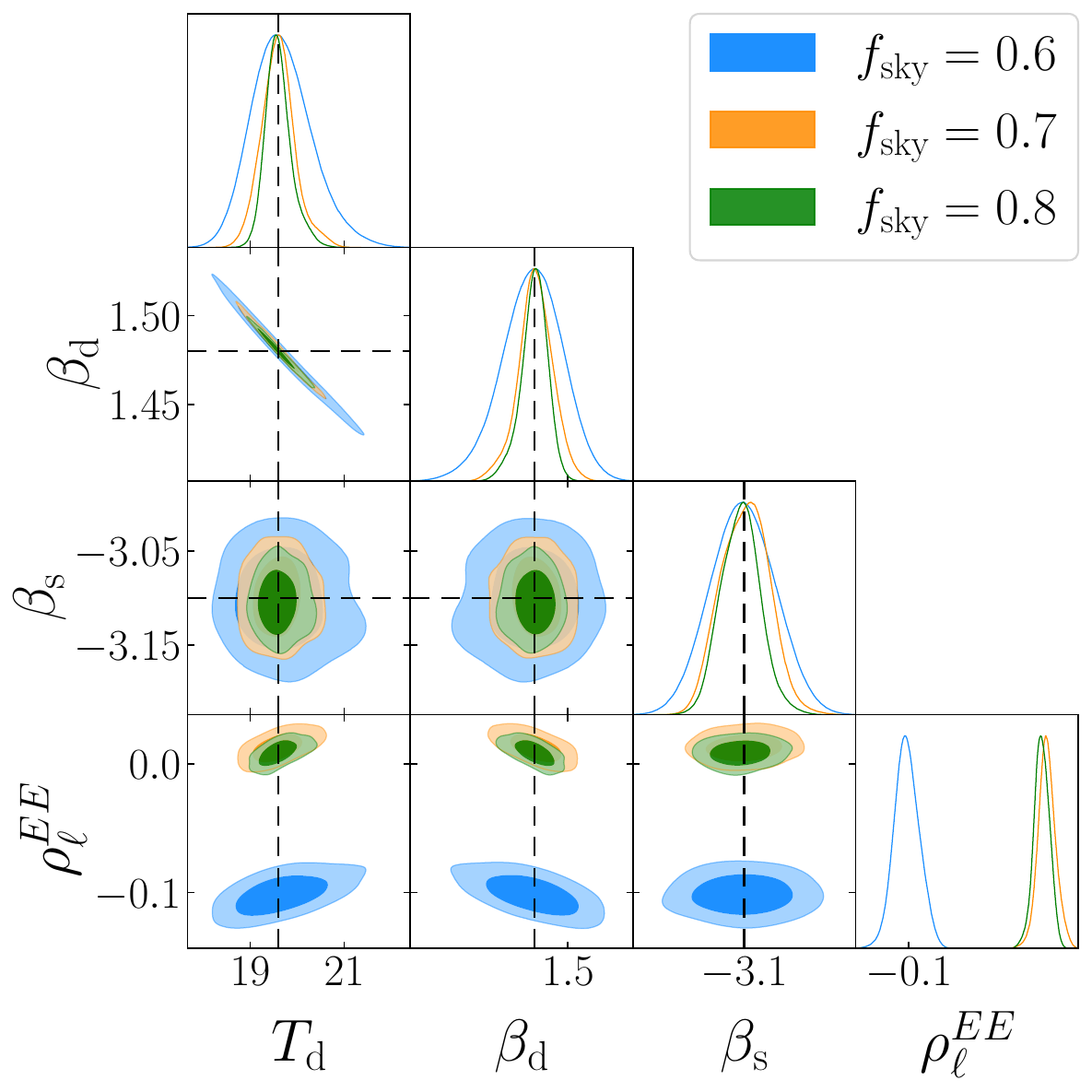}
    \includegraphics[width=0.49\linewidth]{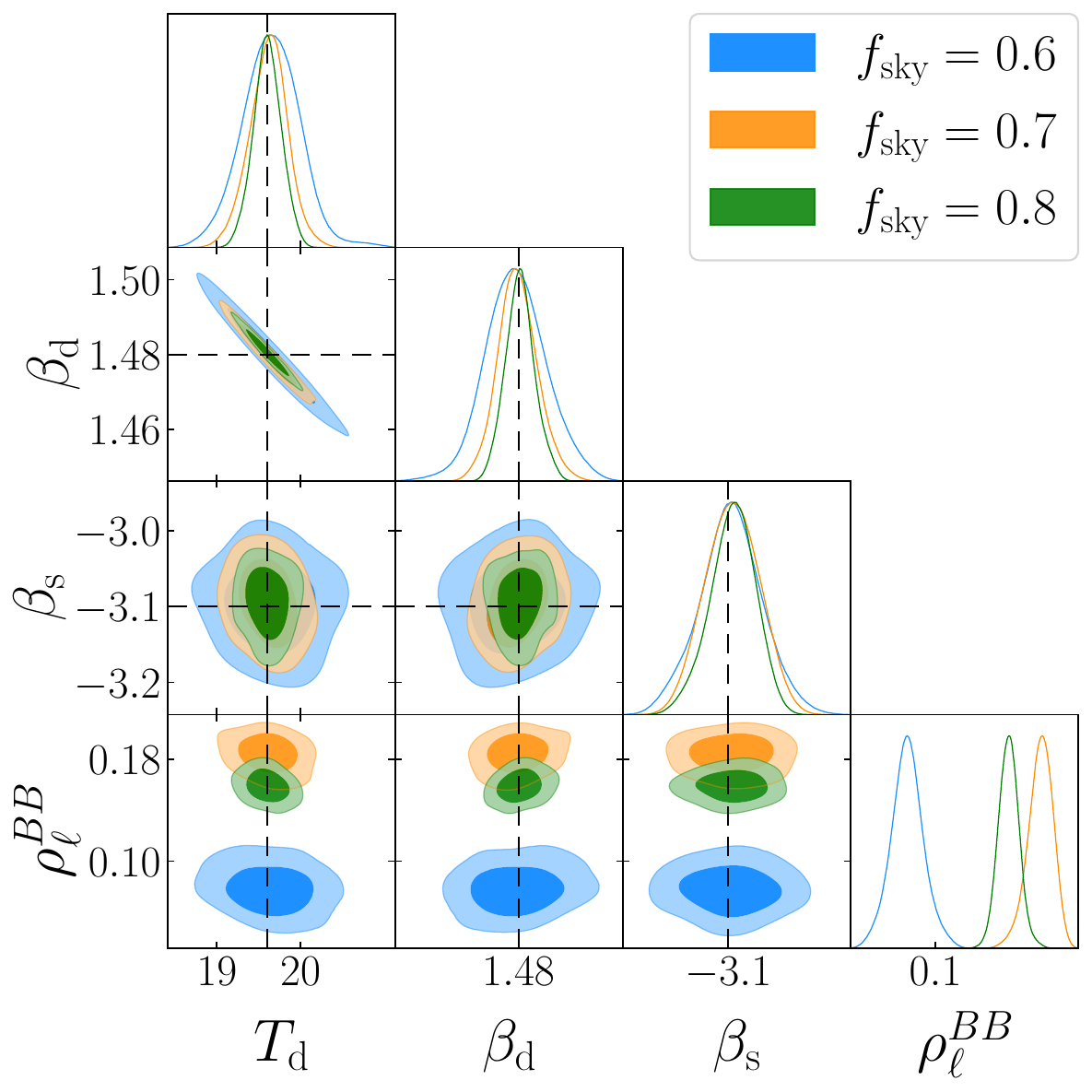}
    \caption{Distribution of parameters $T_{\rm d}$, $\beta_{\rm d}$, $\beta_{\rm s}$, and $\rho$ for $E$-mode (left) and $B$-mode (right) cross-frequency angular power spectra, obtained from the fits of $N_{\rm sim}$ simulations with the \texttt{d9s4} Galactic emission model using the masks presented in section~\ref{sec:masks}. The contours represent \textit{LiteBIRD} $1 \, \sigma$ and $2 \, \sigma$ confidence intervals in the multipole bin centered on $\ell = 26.5$. The input values are shown by the black dashed lines.}
    \label{fig:fsky_comparison}
\end{figure}

We discuss the effect of the sky coverage on the fitted dust and synchrotron parameters, in the simple case of the \texttt{d9s4} Galactic emission model. Figure~\ref{fig:fsky_comparison} shows the distribution obtained for the dust temperature, dust and synchrotron spectral indices and spatial correlation, for the $E$ and $B$ cross-frequency angular power spectra with the three Galactic masks presented in section~\ref{sec:masks}. A higher sky coverage has the effect of increasing the signal-to-noise ratio. Thus, we find as expected that the error bars on the fitted parameters are tightened for the fits with $f_{\rm sky} = 0.8$ and widened for the ones with $f_{\rm sky} = 0.6$, with respect to the baseline of this article. \par

\begin{figure}[t]
    \centering
    \includegraphics[width=\linewidth]{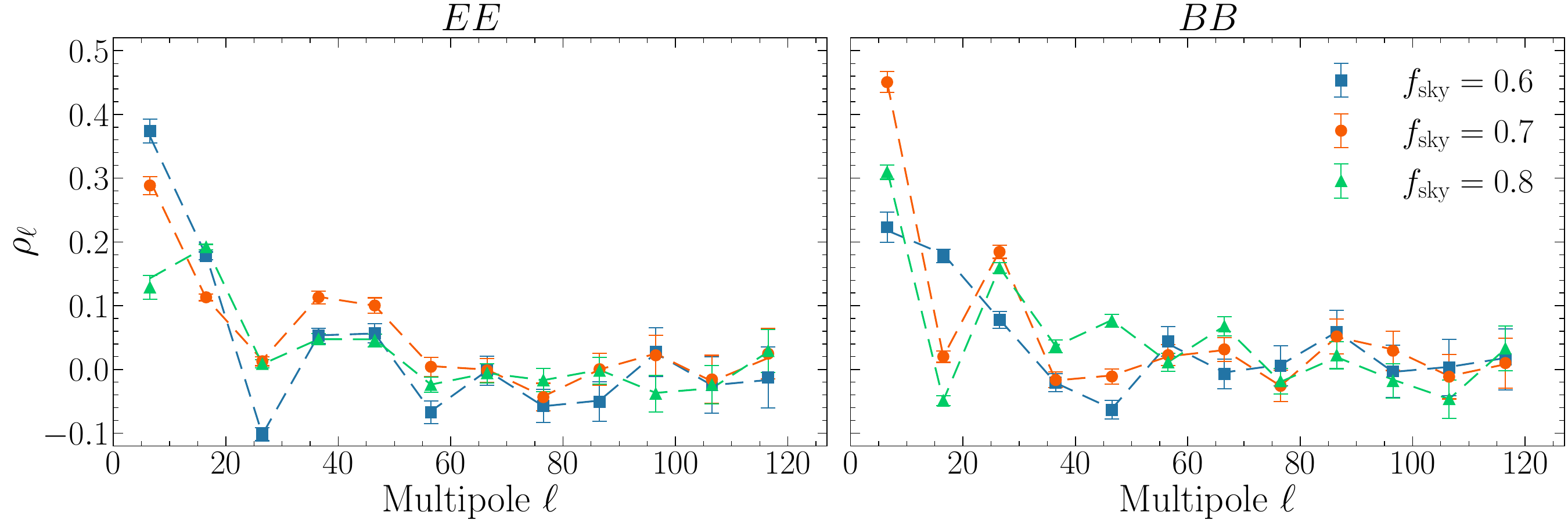}
    \caption{Dust and synchrotron spatial correlation inferred from the fits of the $E$-mode (left) and $B$-mode (right) cross-frequency angular power spectra, for the different Galactic masks introduced in section~\ref{sec:masks}. The considered Galactic emission model is \texttt{d9s4}, and the corresponding input coefficients for each sky coverage are represented by the colored dashed lines.}
    \label{fig:fsky_comparison_rho}
\end{figure}

Additionally, the best-fit values of the correlation coefficient between dust and synchrotron emission, $\rho_\ell^{XX}$, is highly dependent on the chosen mask. Indeed, regions of the sky closer to the Galactic plane are probed as the sky coverage is increased, and we expect that at large $f_{\rm sky}$, statistics of Galactic emission is dominated by low Galactic latitudes. $\rho_\ell^{XX}$ also fluctuates from one multipole bin to another, see figure~\ref{fig:fsky_comparison_rho} illustrating these two effects for the Galactic masks considered in this study. These results are consistent with what is observed in ref.~\cite{Planck2018_XI}, the \texttt{PySM} thermal dust and synchrotron radiation models being the result of the \textit{Planck} 2018 data analysis.

\section{\boldmath Fits to the $E$-mode power spectra} \label{app:E_figures}

\begin{figure}[t]
    \centering
    \includegraphics[width=1\linewidth]{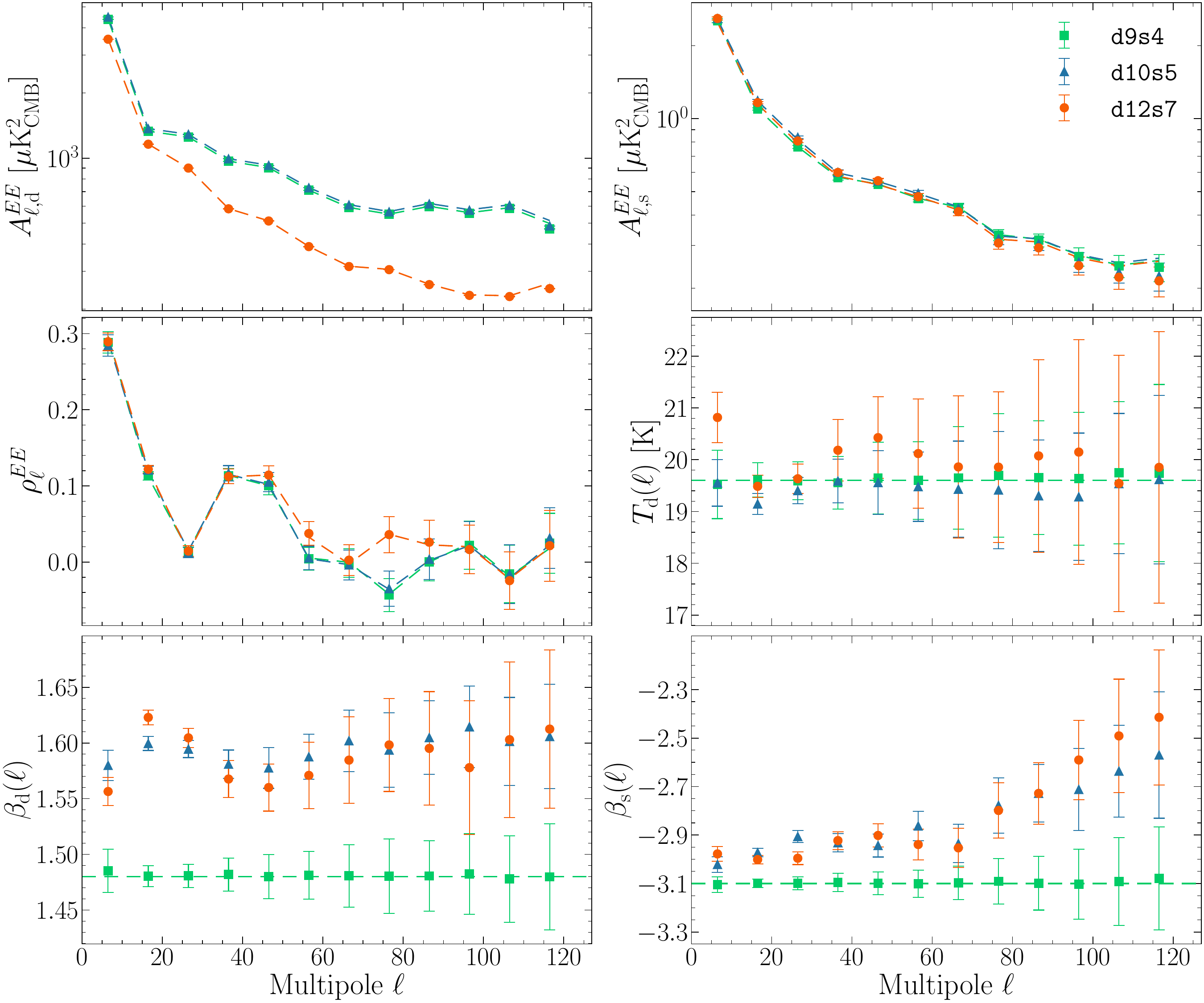}
    \caption{Same as figure~\ref{fig:models}, but for the $E$-mode cross-frequency angular power spectra.}
    \label{fig:models_EE}
\end{figure}

\begin{figure}[t]
    \centering
    \includegraphics[width=0.5\linewidth]{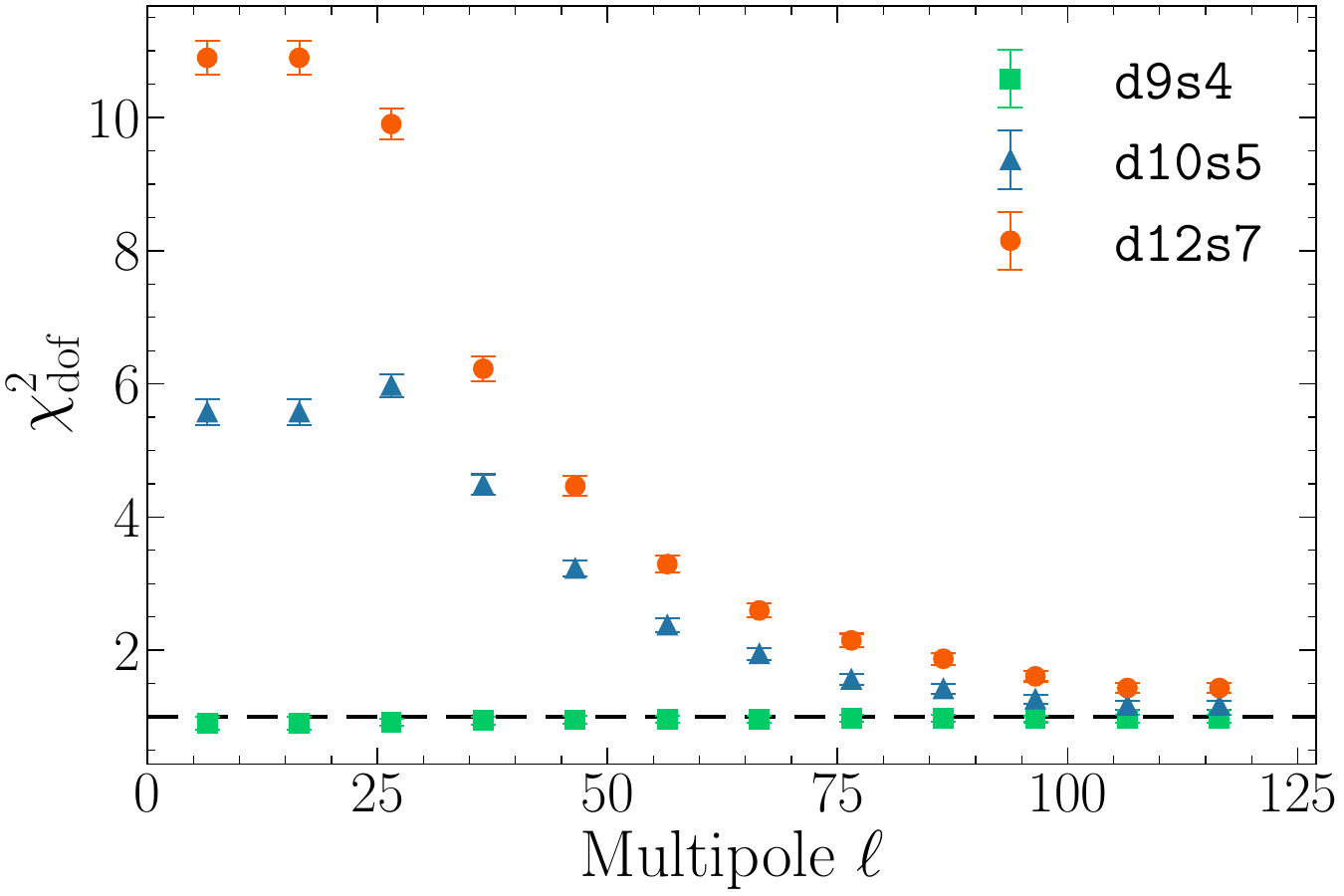}
    \caption{Same as figure~\ref{fig:chi2r}, but for the $E$-mode cross-frequency angular power spectra.}
    \label{fig:chi2r_EE}
\end{figure}

\begin{figure}[t]
    \centering
    \includegraphics[width=\linewidth]{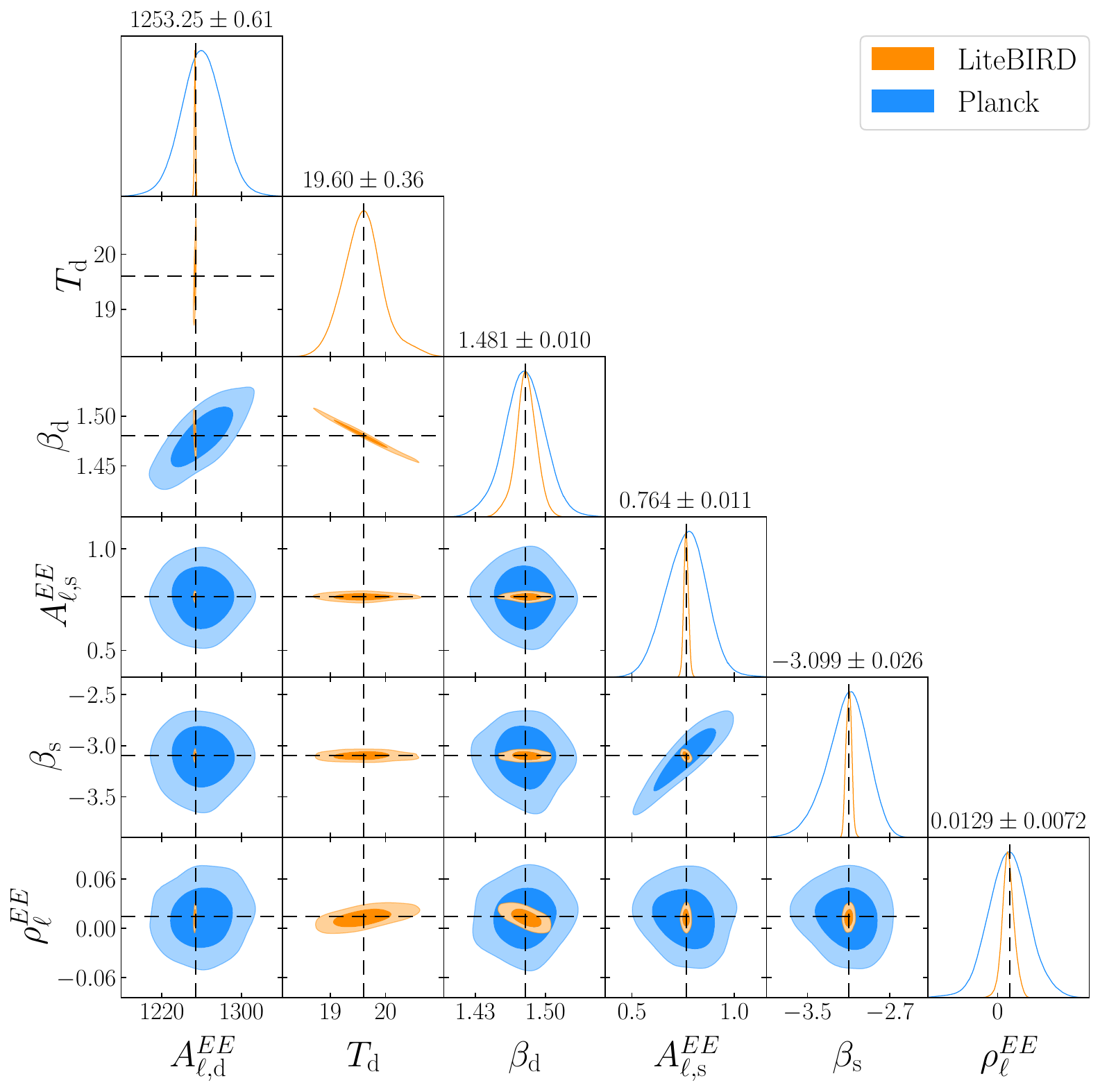}
    \caption{Same as figure~\ref{fig:PlvsLB}, but for the $E$-mode cross-frequency angular power spectra.}
    \label{fig:PlvsLB_EE}
\end{figure}

Figures~\ref{fig:models_EE}, \ref{fig:chi2r_EE} and \ref{fig:PlvsLB_EE} show the equivalents of figures~\ref{fig:models}, \ref{fig:chi2r} and \ref{fig:PlvsLB} for the $EE$ cross-frequency angular power spectra, instead of the $B$-mode power spectra as presented in section~\ref{sec:model_comparison}.

\section{Fits to \textit{Planck} simulations} \label{app:planck_figures}

\begin{figure}[t]
    \centering
    \includegraphics[width=\linewidth]{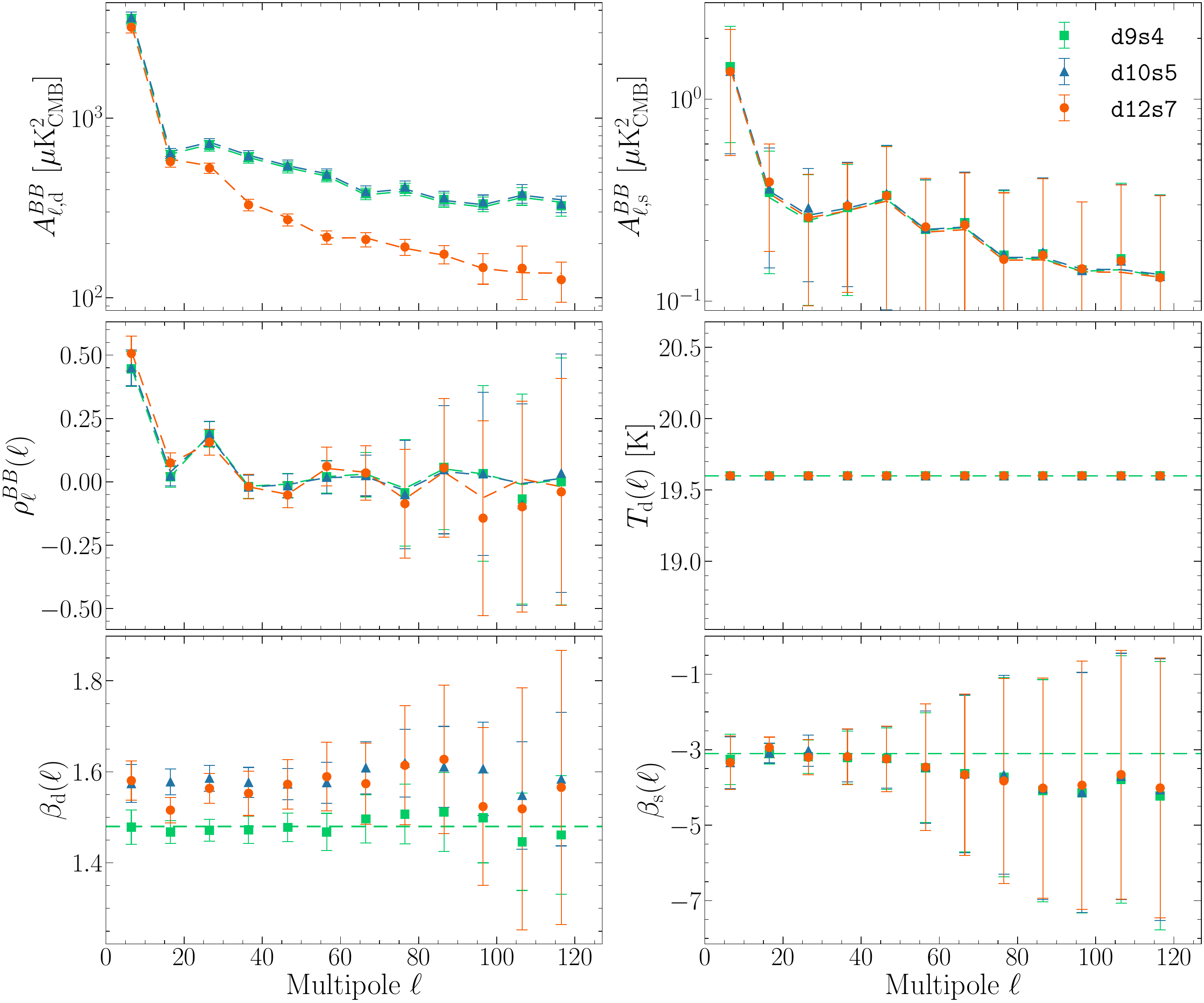}
    \caption{Same as figure~\ref{fig:models}, but for the simulations of \textit{Planck} frequency channels. For the three Galactic emission models, the dust MBB temperature is fixed to its fiducial value, $T_{\rm d}(\ell) = 19.6$~K.}
    \label{fig:models_planck}
\end{figure}

\begin{figure}[t]
    \centering
    \includegraphics[width=0.5\linewidth]{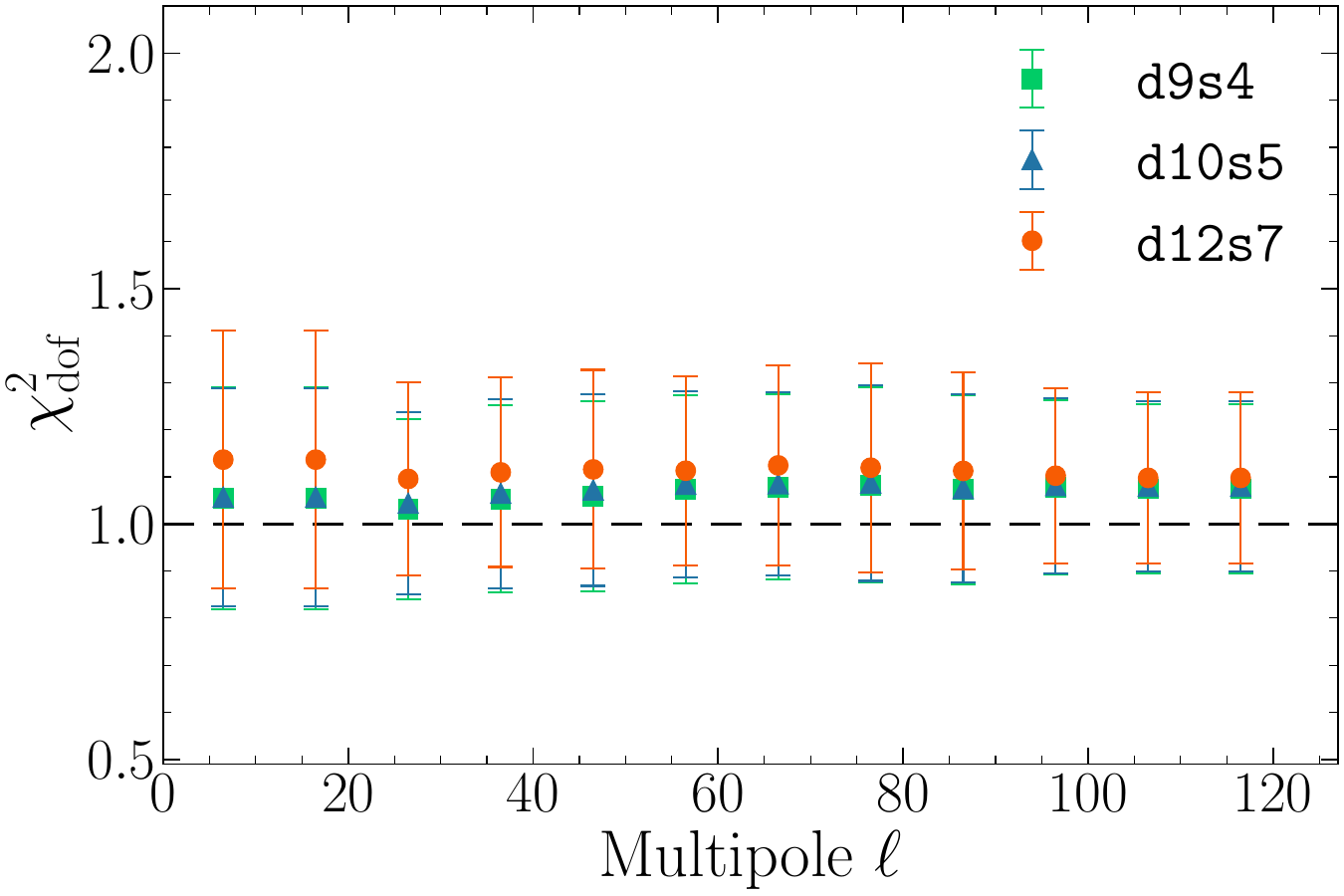}
    \caption{Same as figure~\ref{fig:chi2r}, but for the simulations of \textit{Planck} frequency channels.}
    \label{fig:chi2r_planck}
\end{figure}

\begin{figure}[t]
    \centering
    \includegraphics[width=\linewidth]{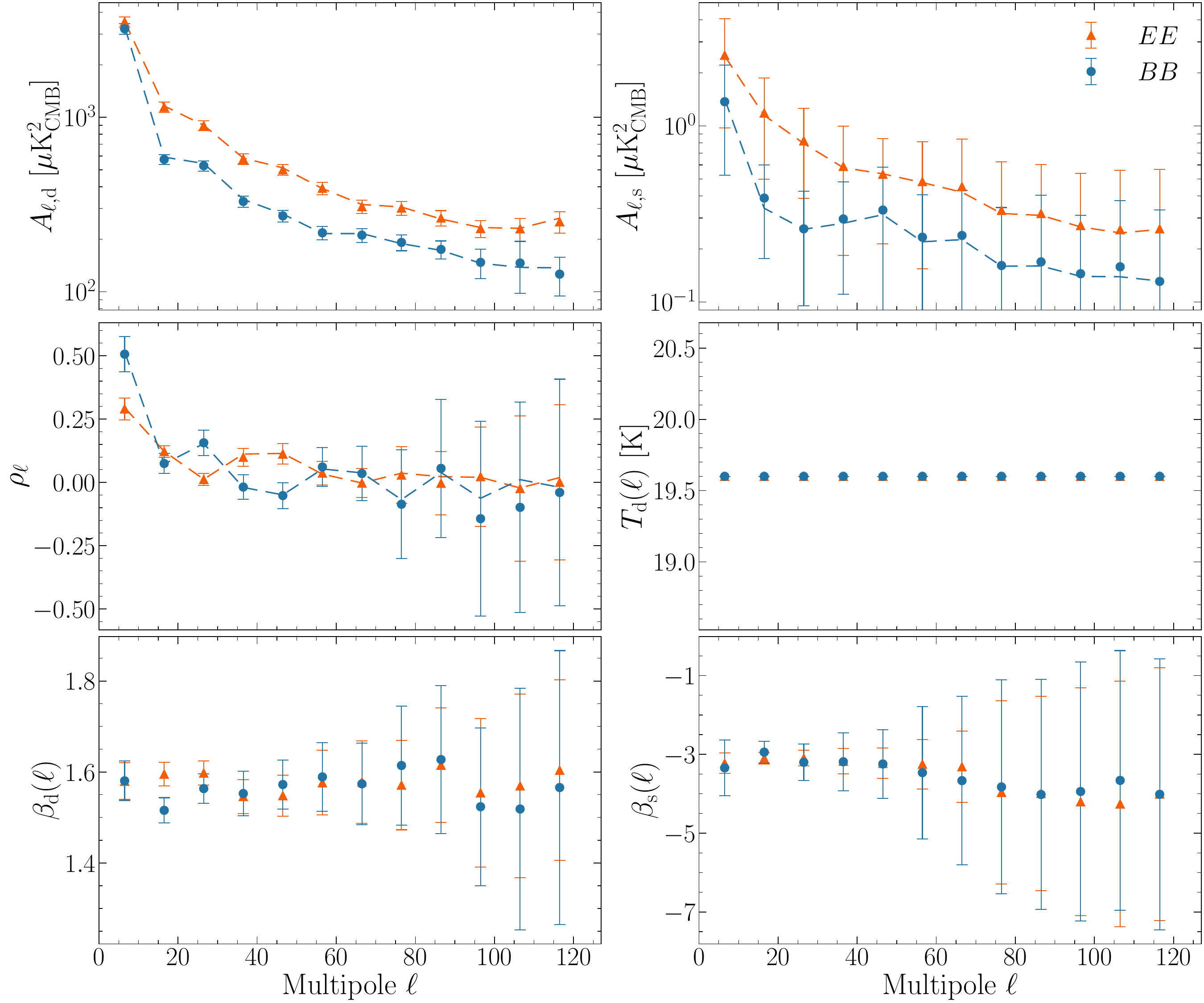}
    \caption{Same as figure~\ref{fig:modes}, but for the simulations of \textit{Planck} frequency channels. The dust MBB temperature is fixed to its fiducial value, $T_{\rm d}(\ell) = 19.6$~K.}
    \label{fig:modes_planck}
\end{figure}

In this appendix are shown the fits to the \textit{Planck} simulations, along with comparisons between the different Galactic emission models in $E$ modes and $B$ modes. The equivalents of figures~\ref{fig:models}, \ref{fig:chi2r} and \ref{fig:modes} are shown in figures~\ref{fig:models_planck}, \ref{fig:chi2r_planck} and \ref{fig:modes_planck}, respectively.

\section{Extending the moment expansion to second order} \label{app:high_order_moments}

\begin{figure}[t]
    \centering
    \includegraphics[width=\linewidth]{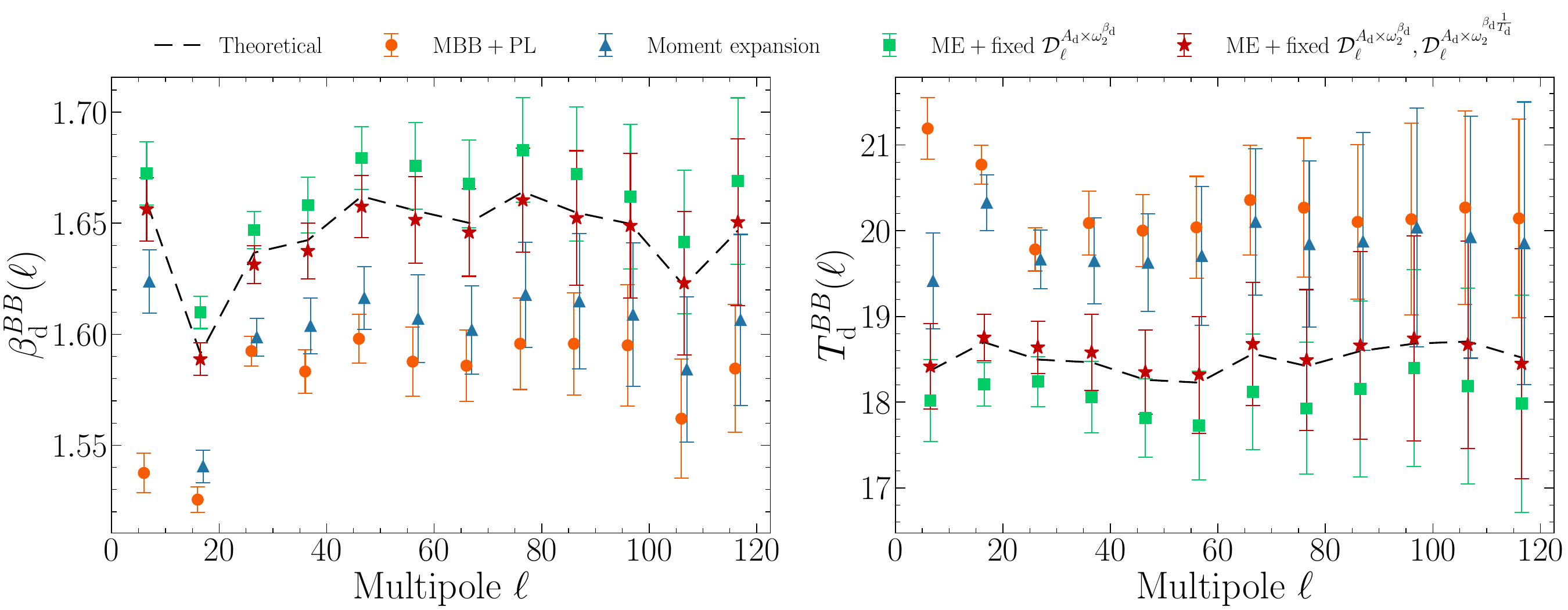}
    \caption{Dust spectral parameters obtained from the fits of the $B$-mode power spectra of mock data with the \texttt{d12s7} foreground model. The dashed lines correspond to theoretical expectations from eq.~\eqref{eq:beta_ell}. We show the results obtained from zeroth-order parametrization as orange circles, and from first-order moment expansion as blue triangles. When fixing the most significant higher-order terms to their input values, we obtain the fits drawn as green squares and red stars, successively.}
    \label{fig:beta_ell}
\end{figure}

As discussed in section~\ref{sec:mom}, moment expansion at first order is not sufficient to recover unbiased mean values of the spectral parameters as a function of $\ell$ for the most complex foreground models (see figure~\ref{fig:beta_ell} for \texttt{d12s7} in $B$ modes), hence the need to include higher-order terms. By adding successively the most significant second-order moments to the fitted model, we verified that the ones having an effect on the recovered values of $\Pi(\ell)$ are actually those that strongly correlate with $\mathcal{D}_\ell^{A \times \omega_1^\Pi}$. Indeed, by not including them in the model, we implicitly fix their values to zero. If their true values differ from that assumption, first-order moments will try to compensate for this mismatch, affecting the value of $\Pi(\ell)$ through eq.~\eqref{eq:beta_ell}. However, adding these second-order terms to the model is not an easy task: because of the degeneracies between parameters, their error bars increase by several orders of magnitude, making the fits numerically unstable. Therefore, to illustrate our last statement, we re-ran our fits by fixing the second-order moments that correlate with $\mathcal{D}_\ell^{A \times \omega_1^\Pi}$ to their theoretical values computed from the input spectral parameters maps, adding successively the ones that contribute the most to the signal. The results are shown in figure~\ref{fig:beta_ell}, where we clearly see the benefit of fixing the values of the dominant second-order coefficients. Such an approach is of course unrealistic and is possible here only because we know the inputs of the simulations. Investigations on more realistic situations are left for future work. \par

\end{document}